\documentclass[fleqn,usenatbib]{mnras}

\usepackage{newtxtext,newtxmath}
\usepackage[T1]{fontenc}

\DeclareRobustCommand{\VAN}[3]{#2}
\let\VANthebibliography\thebibliography
\def\thebibliography{\DeclareRobustCommand{\VAN}[3]{##3}\VANthebibliography}

\usepackage{graphicx}	% Including Figure files
\usepackage{amsmath}	% Advanced maths commands
\usepackage{xcolor}
\usepackage{multirow}
\usepackage{soul}

\definecolor{mlp}{rgb}{0.294, 0.612, 0.827}

\newcommand{\ms}{\mathrm{m\,s^{-1}}}
\newcommand{\cms}{\mathrm{cm\,s^{-1}}}

\title[Towards understanding stellar variability]{Towards understanding stellar variability at the sub m/s level: granulation-induced variability across the optical spectrum}

\author[C Lagae et al.]{\parbox{\textwidth}{\Large
Cis Lagae,$^{1, 2}$\thanks{E-mail: cis.lagae@warwick.ac.uk}
Ginger Frame,$^{1, 2}$
Heather M. Cegla,$^{1, 2}$
Veronika Witzke,$^{3}$
René Holzreuter,$^{4}$
Sergiy Shelyag,$^{5}$
Christopher Watson,$^{6}$
Alexander Shapiro,$^{3,4}$
}
\vspace{0.2cm}
\\
$^{1}$Department of Physics, University of Warwick, Gibbet Hill Road, Coventry CV4 7AL, UK\\
$^{2}$Centre for Exoplanets and Habitability, University of Warwick, Gibbet Hill Road, Coventry CV4 7AL, UK\\
$^{3}$ University of Graz, Institute of Physics, Universit\"atsplatz 5, 8010 Graz, Austria\\
$^{4}$ Max-Planck-Institut f\"ur Sonnensystemforschung, Justus-von-Liebig-Weg 3, 37077 G\"ottingen, Germany\\
$^{5}$College of Science and Engineering, Flinders University, Tonsley Innovation District, 5042, South Australia, Australia.\\
$^{6}$Astrophysics Research Centre, School of Mathematics and Physics, Queen’s University Belfast, Belfast, BT7 1NN, UK \\
}

\date{ Accepted 2026 July 21. Received 2026 July 21; in original form 2026 April 14}

\pubyear{\the\year{}}

\begin{document}
\label{firstpage}
\pagerange{\pageref{firstpage}--\pageref{lastpage}}
\maketitle

\begin{abstract}
Detecting the radial velocity signal of Earth-mass exoplanets requires the characterisation and removal of granulation-induced radial velocity variability from spectroscopic observations. By coupling state-of-the-art three-dimensional (3D) hydrodynamic (HD) simulations to a radiative transfer code, we can isolate and study the effect of granulation on stellar lines. In this study we isolated the impact of granulation on spectral line shapes and shifts for the largest and most diverse synthetic spectral line sample to date. Our aims were twofold. First, we quantified how granulation affects the temporal evolution of shapes and shifts of 72 unblended spectral lines in two wavelength regions: $5500-5600$ \AA~and $6100-6200$ \AA, from disc centre to the stellar limb. Second, we investigated if spectral lines behave coherently in their line shape variability and, if so, can lines be grouped together that behave similarly. We find that weak lines show the largest radial-velocity variability due to granulation, up to $40~\ms$ at disc centre and $50~\ms$ at the stellar limb. On the other hand, strong lines exhibit larger variability in equivalent width than weak lines across the stellar disc. In addition, the equivalent width and line depth of a spectral line are strongly linearly correlated with its radial velocity. While granulation affects all three of these quantities, planet-induced Doppler shifts only affect radial velocities, opening the door to new granulation-mitigation methods. Lastly, we find that the radial velocity and equivalent width of most spectral lines in our sample evolve coherently in time, with the \ion{Fe}{I} and \ion{Ca}{I} lines behaving the most similarly. Due to the coherency,  line blends will not significantly affect the temporal behaviour of RV and line shape, as induced by granulation. Besides, the coherency between spectral lines offer the opportunity to create disc-integrated spectra of many lines simultaneously, which will be explored in future work.

\end{abstract}

% Select between one and six entries from the list of -traceback approved keywords.
\begin{keywords}
techniques: radial velocities -- Sun: granulation -- line: profiles -- hydrodynamics -- stars: solar-type -- methods: analytical
% stars: activity -- stars: atmospheres -- stars: solar-type
\end{keywords}

%%%%%%%%%%%%%%%%%%%%%%%%%%%%%%%%%%%%%%%%%%%%%%%%%%

%%%%%%%%%%%%%%%%% BODY OF PAPER %%%%%%%%%%%%%%%%%%

\section{Introduction}

Finding exoplanets similar to our own Earth remains one of the main goals of the exoplanet field. The upcoming space mission PLATO \citep{Rauer14,Rauer25} and the ground-based survey the Terra Hunting Experiment \citep{Thompson16} are expected to detect the first Earth-analogues. However, these detections will require confirmation and characterisation, for example, by constraining the planetary mass with the radial-velocity (RV) technique. The RV signal of an Earth-analogue orbiting a Sun-like star is $\sim 0.1~\ms$, which is technically within the precision of current high-resolution spectrographs such as \texttt{ESPRESSO} \citep{Pepe13}, with other spectrographs such as \texttt{EXPRES} \citep{expres,Blackman20} and \texttt{NEID} \citep{Schwab16} reaching $\sim 0.3~\ms$. However, stellar variability originating from the host star introduces RV variations that can completely obscure such planetary signals. This variability originates from the stellar surface and includes a wide variety of phenomena operating on different timescales. These range from stellar p-modes (minutes), surface granulation (minutes to hours), supergranulation (hours to days) and magnetic activity such as starspots (days to months). All these sources of stellar activity impact the formation of spectral lines in different ways, changing their strength, shape and introducing RV shifts. For example, solar surface granulation induces RV noise of the order $0.2-0.8~\ms$ \citep{Elsworth1994, Meunier2015, Cameron2019, Sulis20, dalal2023, Lakeland2024, Frame26}. In addition, different sources of stellar RV noise do not act independently of one another. For example, increasing the global magnetic field strength leads to increased suppression of surface convection which, in turn, affects the formation of spectral lines. Many techniques have been developed to remove as much of the stellar noise as possible, but no method exists yet that consistently reduces the observed RV root-mean-square (rms) variability to sub-meter-per-second levels \citep{zhao2022}. Specifically, removing the RV variability from surface granulation, at various magnetic field strengths, remains an open problem.

Significant effort has been undertaken to characterise how granulation affects the shape and shifts of spectral lines across the optical spectrum, both from a theoretical and empirical direction. Stellar surface granulation is defined as the mix of broad hot upflows (granules) bounded by narrow cool downdrafts (intergranular lanes). Spectral lines forming above these features will vary in strength, due to the difference in temperature structure, and will either be blue-or redshifted. Since the granules are brighter and occupy a larger surface area than intergranular lanes, and since the photospheric velocity field decreases with increasing altitude, \cite{Bray1978b} and \cite{Dravins1981} correctly attributed surface granulation as the cause of the observed blueshift and `C'-shape of bisectors from disc-integrated spectral lines. Extensive follow-up studies concluded that the convective blueshift of spectral lines is strongly correlated with their line depth \citep{AllendePrieto1998,Asplund00c,Ramirez08,Dravins2008,Gray09,Reiners16,Liebing21,Liebing23,Ellwarth23}. The weakest lines, forming deeper in the atmosphere, have blueshifts of the order of $~500~\ms$, while stronger lines, that form higher up in the atmosphere, show increasingly smaller blueshifts and even redshifts. Similarly, looking at disc-resolved solar observations, \cite{lohner2019} and \cite{Ellwarth23} found that convective blueshifts decrease in strength from disc-centre to the stellar limb. These findings have also been recovered theoretically, e.g. by \cite{Beeck2015} and \cite{cegla2018}, who synthesised spectral line profiles emergent from solar atmospheres simulated with the 3D radiative-MHD code MURaM \citep{Vogler2005}. Recently, \cite{Palumbo24} has investigated the centre-to-limb variation (CLV) of line shifts for magnetically active regions using observational data from the Solar Dynamics Observatory (SDO). They find that the CLV of the quiet Sun is qualitatively similar to earlier aforementioned work, while the CLV's of plages and magnetic network are generally more redshifted with higher velocity shifts near disc centre.

For exoplanet hunting purposes, characterising the temporal variability of the aforementioned granulation-induced line shifts and shapes is arguably more important. Specifically, it is of interest how granulation affects spectral lines differentially and if there are any dependencies on atomic parameters or line strength. Empirically, \cite{John25} investigated how the observed RV's for a K-dwarf vary with spectral line depth, finding that shallower lines exhibit larger RV variability than deeper lines. In their analysis, they made sure to bin the RV observations accordingly to minimise the influence of p-modes, which introduce RV variations of the order $2~\ms$ on timescales of $\sim5$ min (for the Sun). A similar, albeit, weak relation between line depth and RV has been found by \cite{grass1}, who developed a simulator, based on archival spatially resolved solar spectra, to empirically model the impact of granulation on spectral lines. From a theoretical viewpoint, \cite{Dravins2023} and \cite{Sowmya26} created synthetic spectra of the Sun, computed from three dimensional (3D) hydrodynamic (HD) and magneto-hydrodynamic (MHD) simulations of the solar surface, respectively. Using these models, they investigated the sensitivity of different spectral lines to the solar surface structure. They concluded that the RV rms of individual lines are correlated with lower excitation potential, oscillator strength, line depth and wavelength. One important difference compared to the work of \cite{Dravins2023} is that the methods used by \cite{Sowmya26}, and this work, allow to avoid contamination by the p-modes, essentially isolating the granulation effect.

The main proposed solution to reduce the granulation RV noise is by binning multiple nightly observations over several days. For example, \cite{Dumusque2011} tested different observing strategies on generated synthetic RV observations, based on existing HARPS asteroseismology measurements of subgiant and dwarf stars. They found that they could achieve a $30~\cms$ precision when binning three nightly exposures of 10 mins, separated by 2 hours, for ten consecutive nights. Similarly, \cite{Meunier2015} created a toy model to simulate solar surface granulation and its resulting spectral signature, to find an optimal observing strategy. They concluded that reaching $10~\cms$ precision is possible when binning five separate exposures of 26 min each over six to ten consecutive nights. Both studies agree that binning a single entire night of observational time is insufficient to reach $10~\cms$ precision. 

Recent work has taken a more physically motivated approach to tackle the granulation-induced RV variability problem. Both the paper series by Cegla and collaborators \citep{cegla2013,cegla2018,cegla2019,frame2025,Frame26}, and Palumbo and collaborators \citep{grass1,grass2} who presented a new method to generate disc-integrated spectra to study how granulation impacts spectral line formation. Specifically, the parametrisation method from \cite{cegla2013,cegla2018} utilises synthetic spectra from 3D solar MHD model atmospheres to remove unwanted p-modes and create disc-integrated spectral line profiles. Their work was inspired by an earlier study from \cite{Dravins1990}, who first proposed to classify spatially resolved spectral lines into distinct groups. \cite{cegla2018} applied this method to synthesise disc-integrated spectra for the \ion{Fe}{I} 6302 \AA~line using a Solar MHD model with a vertical background magnetic field of $B=200~$G. In a follow-up study, \cite{cegla2019} found that measurements of the spectral line equivalent width and bisector shape vary linearly with granulation RV, enabling a reduction in RV rms of up to $60\%$ in the best case. Complementary to the work of \cite{cegla2019}, the GRanulation And Spectrum Simulator (\texttt{GRASS}) of \cite{grass1} utilises disc-resolved solar observations to create new synthetic disc-integrated line profiles. Using \texttt{GRASS}, the authors computed 40 min of disc-integrated spectra for 22 spectral lines to study their granulation-induced RV rms. They found that, in the best case, correlations between line bisector diagnostics and RV are able to reduce the RV rms by 25 to 35 \%.  More recently, \cite{frame2025, Frame26} built upon the framework by \cite{cegla2018}, using improved solar HD models ($B=0~$G), and introducing a more rigorous parametrisation method to remove p-modes. They generated disc-integrated spectra for four \ion{Fe}{I} lines using their now publicly available \texttt{DISCO} code\footnote{\url{https://github.com/ginger-frame/DISCO}} \citep{Frame26}. They concluded that line equivalent width and depth show the best linear correlation with granulation RV, leading to a reduction of the RV rms of 60 \% in the best case. Although all previous studies show promising results to minimise granulation noise, they were all produced under ideal circumstances, i.e. high resolution and unrealistic high signal-to-noise (SNR). Both \cite{grass2} and \cite{Frame26} showed that when including photon noise, the correlations between line diagnostics and RV rapidly deteriorate. Using their explored selection of line diagnostics, \cite{Frame26} found that combining information of thousands of lines at $\mathrm{SNR}=1000$ results in a RV rms reduction of at most 10\%.

\begin{figure*}
    \centering
    \includegraphics[width = \textwidth, trim=0cm 0cm 0cm 0cm, clip]{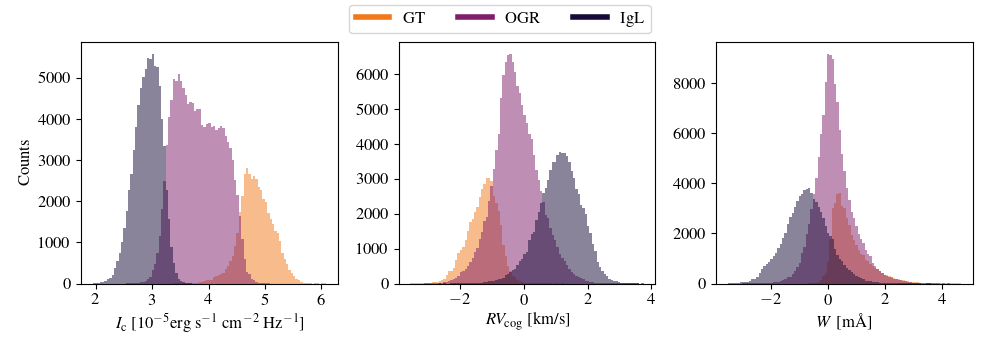}
    \caption{Histograms of the three features: continuum intensity, mean radial velocity and mean equivalent width over all lines, used in the spectral classification process of a single 3D model at disc centre. Each component is colour-coded: granular tops (orange), outer-granular regions (purple), intergranular lanes (black).}
    \label{fig:feature_hist}
\end{figure*}

In this paper, we used the parametrisation technique developed by \cite{cegla2018} and \cite{frame2025}, and expanded it to study how granulation affects the shapes and shifts of $72$ spectral lines across two wavelength regions: $5500-5600$ \AA~and $6100-6200$ \AA. This includes spectral lines from multiple elements, with different excitation and ionisation levels, and oscillator strengths. This is a significant increase in parameter space as compared to previous studies, which typically focused only on a limited sample of spectral lines or elements. The aim to characterise how granulation affects spectral lines is complementary to that of \cite{Sowmya26}, who performed a similar study for a sample of \ion{Fe}{I, II}, and \ion{Ti}{I, II}, lines. However, we take a different approach to remove the p-modes and isolate the granulation signal. \cite{Sowmya26} built on the ergodicity of the solar granulation pattern and used spatial variability of spectral lines across the granulation pattern at a single moment in time to compute the sensitivity of spectral lines to granulation. Here we use the parametrisation technique that allow us to compute not only the sensitivity of the spectral lines but also generate time series of synthetic spectra not affected by p-modes. In Section \ref{sec:method} we summarise the parametrisation technique, detail any changes that were made for this work and validate it to the original method by \cite{frame2025}. In Section \ref{sec:results_ind_lines}, we present how the granulation-induced RV rms varies across our sample of spectral lines, and how it correlates with atomic parameters and line strength. We discuss how it varies from disc centre to the stellar limb, and investigate possible correlations between RV and line strength. Lastly, in Section \ref{sec:discussion} we discuss how our findings can be interpreted in the context of current and future RV observations.

\section{Method}\label{sec:method}

\subsection{Spectrum synthesis}
The solar non-magnetic 3D hydrodynamical (HD) model atmospheres employed in this work were computed by the \texttt{MURaM} code \citep{Vogler2005,witzke2023}. The exact same models were also used in previous work by \cite{frame2025, Frame26}. These models are of box-in-a-star type, set on a uniform mesh of $512\times512\times500$ grid cells, containing on average ten granules at a given time. The physical size of the models is 9 Mm in the horizontal and 5 Mm in the vertical direction, of which 1 Mm lies above the optical surface, with a uniform vertical mesh spacing of $10~\mathrm{km}$. The timeseries consists of 129 snapshots saved at a cadence of 30~s, totalling to 1h~5 min, with a mean effective temperature of $T_\mathrm{eff} \approx 5787~\mathrm{K}$. The surface gravity is fixed at $\log g=4.44$ and chemical composition is taken from \cite{Asplund_2009}.

The spectrum synthesis was performed with an adapted version of \texttt{MPS-ATLAS} \citep{Witzke2021}  that now includes Zeemann-splitting and the ability to compute the four Stokes vectors \citep[][in preparation]{Holzreuter26}, hereafter \texttt{MPS-STOKES-ATLAS}. In this work we limit ourselves to only Stokes I, and leave the study of the other three magnetic Stokes vectors to future work. \texttt{MPS-STOKES-ATLAS} synthesises spectra in 1.5D local thermodynamic equilibrium (LTE), meaning that the spectrum synthesis is performed for each vertical 1D column (also referred to as a pixel) in the 3D model atmosphere independently, but including the local velocity field in all three dimensions. For the spectrum synthesis, the deep layers of the 3D models are trimmed as they do not contribute to the formation of spectral lines. \texttt{MPS-STOKES-ATLAS} assumes solar chemical composition from \cite{Asplund2021}. To compute spectra at limb angles\footnote{Defined as $\mu=\cos \theta$, with $\theta$ the angle between the surface normal and the line of sight.} other than disc centre, the hydrodynamic quantities of the model atmospheres were tilted and interpolated onto slanted rays prior to spectrum synthesis. At small limb angles, such tilting leads to increasingly vertically extended atmospheres, which in turn increases the computational cost when keeping the mesh spacing fixed. Hence, for limb angles smaller than $\mu=1.0$, the hydrodynamic quantities were interpolated on an equidistant mesh in log Rosseland optical depth $-9.04<\tau_\mathrm{ross}<2$ in steps of 0.08, instead of keeping the original geometric $10~\mathrm{km}$ spacing. The resulting tilted atmosphere has 139 grid points in the vertical direction, compared to 150 points for the original model at disc centre. 

In this work, we synthesised spectra for two spectral chunks, each $100$ \AA~wide, at a resolution of $R=500\,000$ and for five positions on the stellar disc: $\mu = 0.2-1.0$ in steps of $\Delta\mu=0.2$. These chunks cover wavelengths (in air units) from $5500-5600$ \AA~and $6100-6200$ \AA, chosen to maximise the number of unblended lines, and to facilitate comparison to previous literature using the \ion{Fe}{I} 6151 and 6173 \AA~lines. The spectral resolution is chosen to be larger than current high-resolution spectrographs, but not too large to constrain the computational cost of spectrum synthesis. For each of these chunks, we created a curated linelist from the VALD3 atomic database \citep{Piskunov1995,Ryabchikova15} such that the remaining spectrum contained only unblended lines. In addition, we cross-matched our curated line-list with the G8 \texttt{ESPRESSO} line mask to make sure to retain lines that are present in the latter. An overview of the linelist with atomic data is given in Appendix \ref{App:AtomicData}.

%This way, we remove unwanted interference from blends on our parametrisation method. 
%In addition, the linelist contains only spectral lines that have ABO values \citep{Anstee1995,Barklem1997} for collisional broadening. 

\section{Parametrisation}

In this paper, we utilised the framework of \cite{frame2025}, hereafter F25, that is tailored to individual spectral lines and expanded it to allow for the parametrisation of larger spectral chunks. The first goal of the parametrisation method is to isolate the spectral signatures of granulation while removing the contribution from p-modes. The second goal is to use this parametrisation to allow for the easy creation of new stellar spectra, free from p-modes and without the need for additional spectrum synthesis. We briefly summarise the method as presented in F25 and then detail the changes made for the current work.

At a specific moment in time, it is the spatial average of bright upflowing granules, dark downdrafts and intermediate regions that leads to the characteristic blueshifted C-shaped line bisector \citep{Dravins1981,frame2025}. Previous studies have shown that it is possible to classify the spectra from these regions into distinct groups \citep{Dravins1990,cegla2013,cegla2018,frame2025}. Here we build on the work by F25 and classify each pixel of the stellar surface into one of the following components: granular tops (GT), outer-granular regions (OGR) and intergranular lanes (IgL). This classification is based on a selection of spectral features which is discussed in the next subsection. Subsequently, spatially averaging all spectra for each component separately, results in three distinct template line profiles. A visual representation of this procedure can be found in \citet[their Figure 5]{frame2025}. In addition, each component has an associated surface filling factor at each point in time. The filling factors are equal to the proportion of pixels classified into each component. By performing this classification for every snapshot of the 3D model in our 1-hour long timeseries, we obtain template line profiles and surface filling factor distributions for each component. By averaging the template line profiles over time, the effect of p-modes is removed \citep{frame2025}. Conversely, the spectral signature of granulation is encoded in the filling factor distributions.

We can then reconstruct each individual line profile $l$, but now void of p-modes, by multiplying the filling factors for each component at each time step $t$ with the spatio-temporally averaged template line profile:
\begin{equation}\label{eq:reconstruction}
    I^l_\mathrm{reconstructed}(t) = f^l_\mathrm{GT}(t)\cdot I^l_\mathrm{GT} + f^l_\mathrm{OGR}(t)\cdot I^l_\mathrm{OGR} + f^l_\mathrm{IgL}(t)\cdot I^l_\mathrm{IgL}~~.
\end{equation}
Here $I^l_\mathrm{GT}$, $I^l_\mathrm{OGR}$ and $I^l_\mathrm{IgL}$ are the template line profiles, and $f^l_\mathrm{GT}$, $f^l_\mathrm{OGR}$ and $f^l_\mathrm{Igl}$ the filling factors of the three components for a single line $l$. These reconstructed spectra are the final data product, and are used to investigate the different effects that granulation has on spectral lines in Section \ref{sec:results_ind_lines}. Besides reconstructing the original spectra, it is also possible to create new spectra using Eq.~\ref{eq:reconstruction}. Instead of using the filling factors of the original timeseries, random filling factors are sampled from the probability distributions and plugged into Eq. \ref{eq:reconstruction} to create new spectra. This sampling procedure is the basis of creating disc-integrated spectra, as it allows to tile a full-disc model of a star with uniquely generated spectra at every spatial position \citep{Frame26}. 

\subsection{Classification}\label{sec:class}
The classification of pixels into one of three components is done through regularised logistic regression. The brightest and faintest $2\%$ continuum intensity pixels on the stellar surface are taken as templates for the supervised machine learning algorithm\footnote{Scikit-learn's LogisticRegression class \citep{scikit-learn}}. Note that the continuum intensity is a direct output from \texttt{MPS-STOKES-ATLAS}, and hence is not measured from the total emergent intensity. The logistic regression algorithm is then used a first time to assign a probability to all other pixels in the 3D model to belong to either the bright (granular tops) or faint group (intergranular lanes), based on user-defined features. From this probability distribution, the third component (outer-granular regions) is defined from the 2\% pixels the logistic regression algorithm is most unsure about. Subsequently, a new multinomial logistic regression model is trained using the three components groups. From this final probability distribution, we can define our three components: granular tops (GT's) are the brighest pixels, intergranular lanes (IgL's) are the faintest, and outergranular regions (OGR's) are the intermediate regions. For a more detailed description, we refer the reader to \cite{frame2025}.

We used two different sets of features in the logistic regression analysis, for two different use cases. The first of which is applied to individual line-by-line analysis, similar to F25, such that each spectral line has its own template line profiles and filling factor distributions. These are used to create the results shown in Section \ref{sec:results_ind_lines}. The second set of features is used in the case where multiple lines are parametrised simultaneously, such that they all share the same filling factor distributions and template spectra. The results of this second use-case are discussed in Section \ref{sec:results_coherence}. 

In the original framework of F25, which considered only individual spectral lines, the features consisted of: RV, line depth and continuum intensity. Since we aimed to extend the method to spectral chunks of $100$ \AA, containing tens of lines, we adapted the features to be more generalisable and computationally cheaper to compute. For the classification of individual lines, we used the following features: RV measured as the first moment of the line profile intensity, equivalent width and mean continuum intensity of the $100$ \AA~chunk. The first moment, or line velocity centre-of-gravity, ($RV_\mathrm{cog}$) of each pixel and line is computed as:
\begin{equation}\label{eq:lambda_cog}
    RV^\mathrm{line}_\mathrm{cog} =\frac{ \int v_D\cdot(1-\frac{I}{I_c})~\mathrm{d}v_D }{ \int (1-\frac{I}{I_c})~\mathrm{d}v_D }~~,
\end{equation}
with $v_D$ the associated Doppler velocity value of each wavelength relative to the line's rest wavelength, $I$ the emergent intensity and $I_c$ the continuum intensity. It is important that the spectral range that is being integrated over in Eq. \ref{eq:lambda_cog} reaches the continuum at both sides of the line profile. Otherwise, the resulting integral will be biased to one side of the line profile. We discuss the potential errors associated with the computation of $RV_\mathrm{cog}$ in Section \ref{sec:RV_method_diff}.
%In addition, the precision of the integral scales with its step size, i.e. spectral resolution, which is not an issue for the high resolution of $R=500~000$ employed in this work.

In Section \ref{sec:results_coherence} we investigated the possibility of classifying all lines in a spectral chunk simultaneously. In this case, we used as features: the mean continuum intensity of the $100$ \AA~chunk, the mean $RV_\mathrm{cog}$ of all lines, and the mean equivalent width of all lines. Specifically, the mean $RV_\mathrm{cog}$, of a single pixel/column in a \texttt{MURaM} snapshot, is defined as the mean of $RV^\mathrm{line}_\mathrm{cog}$ over all lines, where each line RV has been centered by its spatial mean $\langle RV^\mathrm{line}_\mathrm{cog} \rangle$. 
\begin{equation}
    RV_\mathrm{cog} =\frac{1}{N_\mathrm{lines}}\sum_\mathrm{line}\big(RV^\mathrm{line}_\mathrm{cog} - \langle RV^\mathrm{line}_\mathrm{cog} \rangle\big),
\end{equation}
This normalisation ensures that more weight is placed on the strongly spatially varying lines, when computing the mean, instead of the most absolute blueshifted lines. The same averaging procedure and normalisation is also applied to the line equivalent widths, as otherwise the strongest lines will dominate the calculation of the mean. The resulting classification of a single snapshot of the timeseries, using these features, is shown in Figure \ref{fig:feature_hist}. The reconstruction using Eq. \ref{eq:reconstruction} is now performed using the full $100$ \AA~spectral chunk, instead of individual line profiles.

\begin{figure}
    \centering
    \includegraphics[width = \columnwidth]{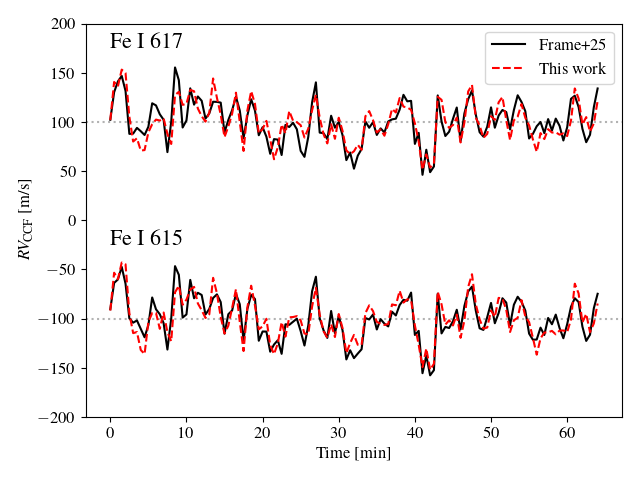}
    \caption[]{RV timeseries of the reconstructed spectra from \cite{frame2025} (solid black) and this work (dashed red), for the \ion{Fe}{I} 6151 and 6173 \AA~lines at disc centre. Both timeseries are shifted by $\pm100~\mathrm{m/s}$ for visualisation purposes.}
    \label{fig:RVtimeseries}
\end{figure}

\begin{figure}
    \centering
    \includegraphics[width = \columnwidth]{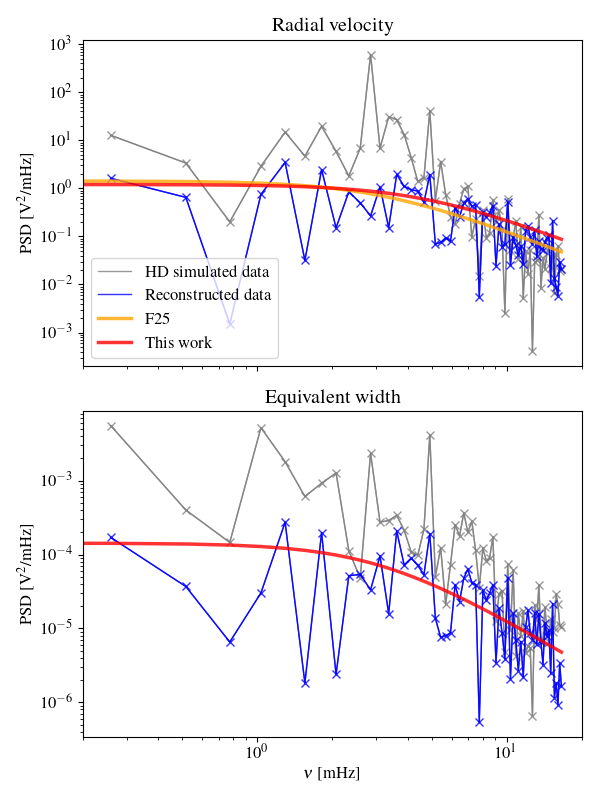}
    \caption[]{Top panel: power spectral density of the RV timeseries of the original (including p-modes, grey) and the reconstructed (only granulation, blue) \ion{Fe}{I} 6173 \AA~line at disc centre. The best joint-fit of the granulation and p-mode components from \cite{frame2025} are overplotted as an orange line. The best fit of the granulation component from this work is shown in red. Bottom panel: power spectral density of the equivalent width timeseries of the same data as the top panel. No joint-fit of the equivalent width was done in \cite{frame2025}. }
    \label{fig:PSD_RV}
\end{figure}

\subsection{Reconstruction and validation}

To validate the parametrisation method, we reconstructed the original spectral timeseries using the newly acquired component template line profiles and filling factors. For each 3D model in the timeseries, we created a reconstructed spectrum of the \ion{Fe}{I} 6151 and 6173 \AA~lines using Eq. \ref{eq:reconstruction}, with their respective filling factors. The temporal mean of the reconstructed disc centre intensity agrees within $0.001\%$ with the original mean spectrum. By comparing the temporal mean, we ensure that all p-modes are averaged out in the original spectra, whereas the reconstructed spectra is free from p-modes by construction. The good agreement between both confirms that our parametrisation and reconstruction is not introducing systematic shifts or offsets in the spectra. 

\begin{figure}
    \centering
    \includegraphics[width = \columnwidth]{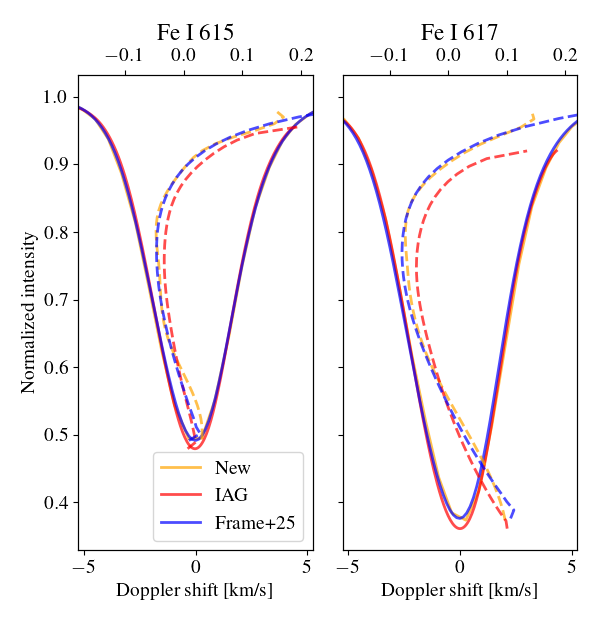}
    \caption[]{Comparison of line bisectors for the \ion{Fe}{I} 6151 and 6173 \AA~lines between \cite{frame2025} (blue), spatially resolved IAG atlas at disc centre (red) and the current work (orange). The spectra from IAG and F25 are downgraded to the resolution used in this work $R=500~000$. The line bisectors (dashed lines) are shifted by their mean value to zero to compare their shape, and are zoomed in for scale, see the separate x-axis at the top.}
    \label{fig:bis_comp}
\end{figure}

The parametrisation method used in this work is equal to that of F25, except in the choice of classification features. In F25, the parametrisation method was validated by comparing the RV power spectral density (VPSD) of the reconstructed spectra to disc-centre observations of the \ion{Fe}{I} 6173 line by the LARS spectrograph \citep{Lohner-Bottcher2017,lohner2019}. Here we validated that our change of classification features has not systematically changed the line properties of the reconstructed spectra, and the corresponding granulation RV variability. We compared the RV timeseries of the reconstructed \ion{Fe}{I} 6151 and \ion{Fe}{I} 6173 lines (Figure \ref{fig:RVtimeseries}), the associated velocity power spectral density (Figure \ref{fig:PSD_RV}), and the line bisectors of the \ion{Fe}{I} 6151 and 6173 lines (Figure \ref{fig:bis_comp}) to those from F25. To match the work of F25, we computed the RV's from Figures \ref{fig:RVtimeseries} $\&$ \ref{fig:PSD_RV} by cross-correlating each spectrum with the temporal mean. The velocity shift was then obtained by upsampling the peak of the cross-correlation function (CCF) to $\Delta v=1$ cm/s using a cubic polynomial. From Figure~\ref{fig:RVtimeseries}, it is seen that our RV's, at disc centre, generally agree with the ones from F25. Comparing the root-mean-square (rms) standard deviation of each timeseries, we find $\sigma_{RV,6151}=21.20~\ms$ and $\sigma_{RV,6173}=20.15~\ms$, compared to F25, $\sigma_{RV,6151}=21.58~\ms$ and $\sigma_{RV,6173}=20.76~\ms$. A similarly good agreement is found at smaller limb angles of $\mu=0.4$: we find $\sigma_{RV,6151}=34.04~\ms$ and $\sigma_{RV,6173}=31.140~\ms$, compared to F25, $\sigma_{RV,6151}=33.90~\ms$ and $\sigma_{RV,6173}=30.37~\ms$. Besides, no systematic differences or trends were found between our RV's and those from F25.

To further compare both, we computed the line velocity centre-of-gravity ($RV_\mathrm{cog}$), RV using the template matching technique \citep{Bouchy01}, see Eq. \ref{eq:Bouchy}, and equivalent width ($W$) for both reconstructed timeseries and lines. The difference in rms between our work and that of F25, at disc centre, is: $\Delta RV_\mathrm{cog} = 0.6~\ms~[2.5\%]$, $\Delta RV_\mathrm{Bouchy} = 0.4~\ms~[3.3\%]$ and $\Delta W = 0.01$~ m\AA~$[9\%]$. The percentages in brackets denote the relative difference to the values of F25. We hypothesise that the difference in equivalent width is due to the sampling in this work being a factor four lower than in F25. This will have a direct effect on the computation of equivalent width, which is performed by numerically integrating over the line profile.

The associated RV power spectral density (PSD) of the \ion{Fe}{I} 6173 line is shown in Figure \ref{fig:PSD_RV}, together with the best joint-fit of the granulation and p-mode component from F25. In their work, a Harvey function \citep{Harvey1985}, sometimes called a super-Lorentzian, was used to fit the granulation component, and a Lorentzian function to fit the p-modes. The Harvey function is defined as:
\begin{equation}
    L_\mathrm{super}(\nu) = \frac{A_{g}}{1 + (2\pi\tau \nu)^\alpha}~~,
\end{equation} 
where \( A_{g} \) is the amplitude, \( \tau \) is the characteristic timescale, and \( \alpha =2\) is the power-law slope. At low frequencies, the Harvey function is flat at a constant value of $A_{g}$. Towards higher frequencies, the power drops to half at a frequency of $1/(2\pi\tau)$ \citep{Kjeldsen2011}. The Lorentzian function is defined as:
\begin{equation}
    L(\nu) = \frac{A_{p} \Gamma^2}{(\nu - \nu_0)^2 + \Gamma^2}, \tag{6}
\end{equation}
with \( A_{p} \) the amplitude, \( \Gamma \) the full-width at half-maximum (FWHM), and \( \nu_0 \) the central frequency. 

In the joint-fit of F25, both the reconstructed and original HD data were fitted simultaneously with the same granulation component, using MCMC sampling, and subsequently compared to disk-centre observations from the LARS spectrograph. Here, we do not aim to redo the same analysis, but instead perform a Levenberg-Marquardt least-squares fit of a Harvey function to the reconstructed data. We find that our best fit parameters for the Harvey function are $A_\mathrm{g} = (1.21\pm0.21)\times10^{-3}~\mathrm{m^2s^{-2}Hz^{-1}}$ and $\tau=34\pm9~\mathrm{s}$ as compared to F25 (their Table A2): $A_\mathrm{g} = (1.43\pm0.8)\times10^{-3}~\mathrm{m^2s^{-2}Hz^{-1}}$ and $\tau=52\pm15~\mathrm{s}$. Our values for the characteristic timescale are similar to those from F25 but do not match perfectly. Therefore, we refit the data from F25 using our fitting method and changing the value of the power-law slope to $\alpha=3.3$, as found by \cite{cegla2018}. Following this, we obtain as best fit parameters for the characteristic timescale: $\tau=28.5\pm5~\mathrm{s}$ for our data, and $\tau=27.3\pm5~\mathrm{s}$ for F25. We also remark that several previous studies have deviated from the original definition of the timescale $\tau$, by for example including the constant $2\pi$ into the timescale, changing its mathematical and physical meaning.  As such, care has to be taken when comparing different studies, and when interpreting the timescale as a physical process.

In addition, we plot the PSD of the equivalent width timeseries following the argument of \cite{cegla2018} that it should behave similarly to the RV's. In conclusion, we find that the reconstructed timeseries of the \ion{Fe}{I} 6151 and 6173 \AA~lines are sufficiently similar to those from F25, providing confidence in our adaption of the parametrisation method. Upcoming spatially resolved observations of the solar surface by the Paranal solar Espresso Telescope \citep[PoET]{Santos25} will enable more detailed validation of our work.

% $\tau=0.22\pm0.05~\mathrm{m Hz^{-1}}$ as compared to F25 (their Table A2): $A_\mathrm{g} = (1.43\pm0.8)\times10^5~\mathrm{m^2s^{-2}Hz^{-1}}$ and $\tau=0.33\pm0.09~\mathrm{m Hz^{-1}}$. In addition, we plot the PSD of the equivalent width timeseries following the argument of \cite{cegla2018} that it should behave similarly to the RV's. In conclusion, we find that the reconstructed timeseries of the \ion{Fe}{I} 6151 and 6173 \AA~lines are sufficiently similar to those from F25, providing confidence in our adaption of the parametrisation method.

In Figure \ref{fig:bis_comp}, we compared the line bisectors of the \ion{Fe}{I} 6151 and 6173 \AA~lines of our reconstructed spectra to the those of \citet[$R_\mathrm{F25}=2~000~000$]{frame2025} and the spatially resolved IAG atlas at disc centre \citep[$R_\mathrm{IAG}\approx700~000$]{IAG}. Both spectral datasets were downgraded to the resolution used in this work, $R=500~000$. The IAG atlas has removed telluric contamination from $\mathrm{H_2O}$ and $\mathrm{O_2}$ lines based on the \texttt{HITRAN2020} database. Excellent agreement is seen between our reconstructed spectral lines and those from F25, with residuals of at most $30~\ms$. If we subtract the mean of each bisector to solely compare the shape, the residuals are at most $5~\ms$, across all limb angles. The strongest deviation occurs near the line core and at the blue-most point of the C-shaped bisector, which will have an impact on certain line diagnostics such as the bisector-inverse-span. Generally, the difference between our work and that of \cite{frame2025} is smaller than the difference compared to the IAG atlas, giving confidence in our adaptation of the parametrisation method. In Appendix \ref{App:bisectors}, we compared the bisectors of the other spectral lines in our two spectral chunks to the IAG atlas. Not all lines are included, as some were too heavily affected by blending in the IAG atlas. The majority of the reconstructed line bisectors behave similarly, at least in part of the line profile, to the IAG atlas. We stress that our spectra only contain pure unblended lines, while the observed solar spectrum does not.

%At mu=0.4 max difference is 5 m/s (after substracting the mean), otherwise diff in mean is 26 m/s
%At mu=0.2 max difference is 4 m/s (after substracting the mean), otherwise diff in mean is 26 m/s
%At mu=0.6 max difference is 2 m/s (after substracting the mean), otherwise diff in mean is 26 m/s

\section{Results}\label{sec:results_ind_lines}
\subsection{Temporal variability of spectral line shift and shape}
% \subsection{Line-by-line granulation signatures}
Since our spectral chunks contain lines from a variety of elements and ionisation stages, with different atomic parameters, we can analyse how granulation affects these differentially. In the following section, we investigated how the granulation-induced temporal variability of RV, equivalent width and line depth depend on lower excitation potential, oscillator strength and line depth. The line shifts and strengths were derived from the reconstructed spectra, where each spectral line has been parametrised and reconstructed individually. The line depth is defined as the distance between the normalised flux at line minimum and the continuum ($=1$), such that strong lines have large values of line depth. Here the normalised flux at line centre is obtained by upsampling the line core using a cubic polynomial and taking the minimum. The computation of RV was done in three different ways: from the $RV_\mathrm{cog}$ as defined in Eq. \ref{eq:lambda_cog}, or determined from the peak of the cross-correlation function (CCF), or using the template matching technique as described in \cite{Bouchy01}. In the CCF method, the reconstructed spectrum at each timestep was cross-correlated with the temporal mean. The velocity shift was then obtained by upsampling the peak of the CCF to $\Delta v=1$ cm/s using a cubic polynomial. In the template matching method, the RV of a single spectral line is derived from a weighted sum of the velocity shifts $\delta V$ at each wavelength point $i$: 
\begin{equation}\label{eq:Bouchy}
    \frac{RV_\mathrm{Bouchy}}{c} = -1 \times \frac{\sum\frac{\delta V(i)}{c}w(i)}{\sum w(i)}~~,
\end{equation}
where the velocity shifts $\delta V$ are computed as:
\begin{equation}
    \frac{\delta V(i)}{c} = \frac{I(i)-I_\mathrm{ref}(i)}{\lambda(i)\big(\partial I_\mathrm{ref}(i)/\partial\lambda(i)\big)}~~,
\end{equation}
and the weights $w$ as:
\begin{equation}
    w(i) = \frac{\lambda^2(i)\big(\partial I_\mathrm{ref}(i)/\partial\lambda(i)\big)^2}{I_\mathrm{ref}(i)}~~.
\end{equation}
Here the spectrum is masked such that the sum covers only a single spectral line, with $I_\mathrm{ref}$ the temporal mean of the intensity. In addition, we added a multiplication with $-1$ in Eq. \ref{eq:Bouchy}, contrary to the original formula by \cite{Bouchy01}, such that blueshifts correspond to negative velocities. Since $RV_\mathrm{Bouchy}$ is computed relative to the temporally averaged spectrum, the resulting line shifts are centred around the mean convective blueshift.

\begin{figure*}
    \centering
    \includegraphics[width = \textwidth, trim=0cm 0cm 0cm 0cm, clip]{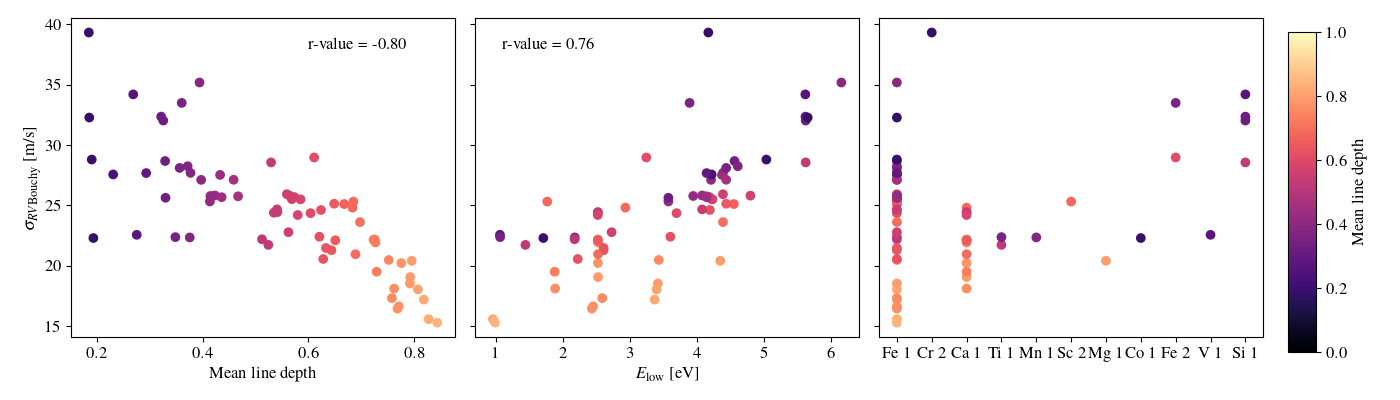}

    \caption{RMS of the temporal evolution of $RV_\mathrm{Bouchy}$ at disc centre, for reconstructed lines in the wavelength region $5500-5600$~\AA~ and $6100-6200$~\AA~. From left to right, $\sigma_{RV, \mathrm{Bouchy}}$ is plotted, and colour coded, against mean line depth, lower excitation potential and element + ionisation stage. The Spearman correlation coefficient (r-value) is shown at the top of the left and middle panel.}
    \label{fig:RV_jitter}
\end{figure*}

\subsubsection*{Radial velocity}\label{sec:results_RV}
To quantify the temporal variability of line depth, equivalent width and RV, we compute their rms standard deviation from the reconstructed timeseries (see as an example Figure \ref{fig:RVtimeseries}). First, in Figure \ref{fig:RV_jitter}, we show the rms variability in $RV_\mathrm{Bouchy}$ at disc centre, with relation to lower excitation potential $E_\mathrm{low}$ and mean line depth. To statistically check for monotonic correlations, we performed a Spearman correlation test. The Spearman rank-order correlation coefficient $r$ ranges between $-1$ and $+1$, with $0$ being no correlation and $\pm1$ implying a monotonic relation. At disc centre, we observe a significant negative trend of $\sigma_{RV_\mathrm{Bouchy}}$ with increasing line depth ($r=-0.8$). Another monotonic correlation can be found between $RV_\mathrm{Bouchy}$ and lower level excitation potential ($r=0.76$), but no monotonic relation is found with the oscillator strength ($r=-0.17$). From Figure \ref{fig:RV_jitter}, the main conclusion is that deep lines exhibit smaller variability in $RV_\mathrm{Bouchy}$ than shallow lines. In addition, the scatter in variability between lines of similar depth is larger for shallow lines, $20~\ms<\sigma_{RV_\mathrm{Bouchy}}<40~\ms$, than for deep lines.

When looking at the solar limb $\mu=0.4$, we find an equally strong relation of $\sigma_{RV_\mathrm{Bouchy}}$ with mean line depth ($r=-0.77$) and a weaker relation with lower excitation potential ($r=0.58$) compared to disc centre. Besides the significance of the monotonic relation, the size of $\sigma_{RV_\mathrm{Bouchy}}$ changes from disc centre to the limb. In general, all lines have larger RV variability at the limb, up to $\sigma_{RV_\mathrm{Bouchy}}<50~\ms$, than at disc-centre. Comparing to literature, \cite{frame2025} found that in their sample of four \ion{Fe}{I} lines, the deep lines exhibit smaller RV variability than shallow lines at all limb angles. \cite{Dravins2023} also found monotonic relations between RV variability and line depth, and lower excitation potential. Although as we discuss in Section \ref{sec:CLV}, the shape of the monotonic relation with line depth changes when p-modes are present. In addition, \cite{Dravins2023} concluded that lines of similar depth exhibit similar RV variability. Looking at Figure \ref{fig:RV_jitter}, this seems to be true for deep lines but breaks down with decreasing line depth, with the most shallow lines showing a large scatter in RV variability. Lastly, using their empirically based spectral simulator \texttt{GRASS}, \cite{grass2} found large differences in variability between different disc-integrated spectral lines, ranging from $45-75~\cms$. However, they did not find any significant underlying correlations with line depth, wavelength, formation temperature or lower excitation potential. 

When instead of $RV_\mathrm{Bouchy}$ we look at $RV_\mathrm{cog}$, we find only a weak correlation between $RV_\mathrm{cog}$ and line depth ($r=-0.43$), a stronger correlation with $E_\mathrm{low}$ ($r=0.75$), and no correlation with $\log gf$ ($r=0.11$). The dependence of $RV_\mathrm{cog}$ variability on increasing lower excitation potential is also recovered in \cite{Sowmya26} for their sample of \ion{Fe}{I} and II lines. Their method utilises the spatial variability in RV of synthetic spectra from a single 3D MHD model of the Sun, as a proxy of the temporal variability. Since no timeseries is used, the temporal influence of p-modes is removed, instead, their spectra are generated at a specific phase of the p-mode oscillations present in the 3D model. Contrary to their work, we do not recover their correlation with oscillator strength. It is difficult to perform a clear one-to-one comparison due to the large difference in methodology. However, this difference should be explored in future work. Already one important distinction is that \cite{Sowmya26} synthesised spectra from a 3D model containing a small-scale-dynamo, compared to the non-magnetic models used in this work. 

Physically, the dependence of $RV_\mathrm{Bouchy}$ variability on line depth can be explained by different lines forming at different heights in the stellar atmosphere. From 3D HD simulations of stellar atmospheres we know that the amplitude of fluctuations in temperature and vertical velocity reaches a maximum below the optical surface \citep{Magic13b, Magic14a}, in the super-adiabatic regime where convection is at its strongest. These variables are shown in Figure \ref{fig:MURAM} in the appendix for the Solar \texttt{MURaM} simulations used in this work. The amplitude of fluctuations in horizontal velocity reaches a maximum slightly above the optical surface. In the line-forming regions, the amplitude of the vertical velocity fluctuations decreases to a minimum at approximately $\log\tau_\mathrm{ross}\approx-2$, whereafter they start increasing again in the higher atmospheric layers. The minimum for temperature fluctuations is reached at $\log\tau_\mathrm{ross}\approx-1$, and for horizontal velocity $\log\tau_\mathrm{ross}\approx-4$. The latter minimum is reached at such high altitudes that you would not expect significant photospheric line formation there. As such, it is expected that shallow lines, which generally form deeper in the atmosphere, show larger variability in line shifts than deep lines, which form at higher altitudes. When looking at the stellar limb, the horizontal velocity fields start dominating and spectral lines are formed at higher altitudes. The same argument as for disc centre holds here, shallow lines form deeper down where the velocity field shows larger fluctuations. 

Besides the physical interpretation, the RV measurement itself will also impact the resulting RV rms. Specifically, the information content of a line differs between deep and shallow lines, and is dependent on the spectral resolution and sampling. \cite{Bouchy01} quantified this information content in the form of a quality factor. They found that a resolution of $R=100\,000$ is sufficient to obtain the highest quality factor, in the majority of cases. At lower spectral resolution, RV measurements can be affected by the difference in information content between deep and shallow lines, and hence lead to larger rms values for weak lines. To confirm that the impact is minimal in our work, we conducted a simple test, comparing the true RV rms of idealised Gaussian line profiles of various line depths, to their measured RV rms. We found that, at the spectral resolution used in this work ($R=500\,000$), the difference in RV rms is negligible and cannot account for the trend seen in Figure \ref{fig:RV_jitter}.

\begin{figure*}
    \centering
    \includegraphics[width = \textwidth, trim=0cm 0cm 0cm 0cm, clip]{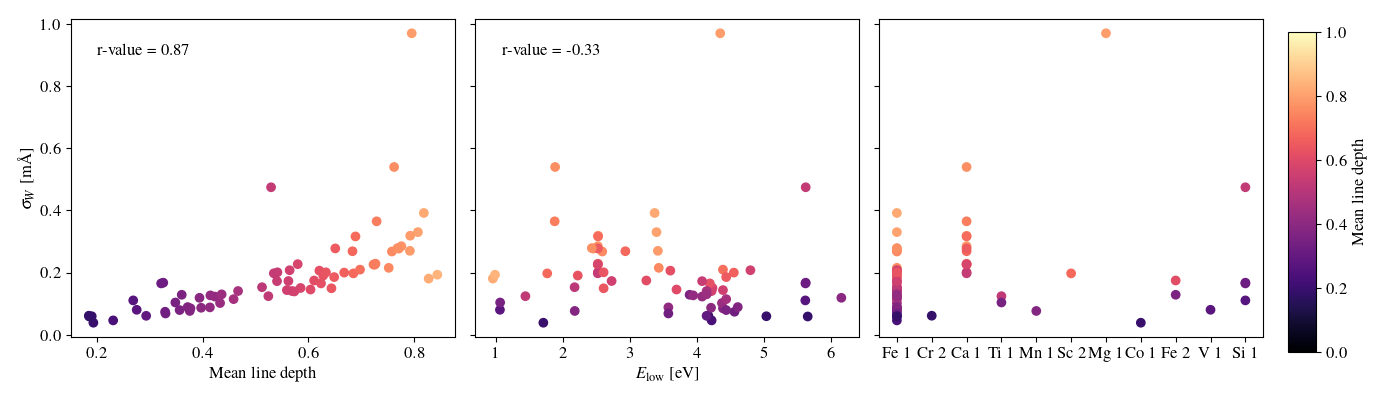}

    \caption{RMS of the temporal evolution of the equivalent width at disc centre, for reconstructed lines in the wavelength region $5500-5600$~\AA~ and $6100-6200$~\AA~. From left to right, equivalent width ($W$) is plotted, and colour coded, against mean line depth, lower excitation potential and element + ionisation stage. The Spearman correlation coefficient (r-value) is shown at the top of the left and middle panel.}
    \label{fig:EW_jitter}
\end{figure*}

\subsubsection*{Equivalent width}
The temporal variability of equivalent width at disc centre is shown in Figure \ref{fig:EW_jitter}. We find a significant monotonic relation with mean line depth ($r=0.87$), similar to $RV_\mathrm{Bouchy}$. Deeper lines have smaller $RV_\mathrm{Bouchy}$ variability, but a larger variability in equivalent width. On the other hand, shallow lines have larger $RV_\mathrm{Bouchy}$ variability, but barely vary in equivalent width. The strong dependency of equivalent width variability on the mean line depth can be explained by the fact that these lines lie on the saturated part of the curve of growth, defined roughly as $W/\lambda_0\gtrapprox-5$\footnote{The transition from the weak part of the curve-of-growth to the saturated part is element and line dependent. The cut-off presented here is indicative for \ion{Fe}{I} \citep{Rutten1983,Tielens1999}.}. The wings of saturated lines are known to be more sensitive to the local velocity field \citep[chapter 17]{Gray_2005}, whereas the line depth becomes less sensitive. In Figure \ref{fig:EW_jitter}, all elements behave similarly except for \ion{Mg}{1}, which exhibits significantly larger variability in equivalent width compared to all other elements, neutral or ionised. Although seemingly an outlier when plotted against mean line depth, the \ion{Mg}{1} line has the largest equivalent width of our sample, $W=370~$m\AA, compared to the second strongest line with $236$ m\AA. By plotting $\sigma_W$ with respect to reduced equivalent width [$W/\lambda_0$], see Figure \ref{fig:tempvar_REW} in the Appendix, we find again a strong monotonic relation, but now without outliers, explaining the large variability for the \ion{Mg}{1} line. At the stellar limb ($\mu=0.4$), the correlation of equivalent width variability with line depth remains ($r=0.79$), but there is more scatter between spectral lines with similar line depth than at disc centre. In addition, there is no correlation with lower excitation potential at the stellar limb ($r=-0.14$). These results are in line with those from \cite{Sowmya26}, who did not find a significant relation between variability in equivalent width and lower excitation potential.

\subsubsection*{Line depth}
\begin{figure*}
    \centering
    \includegraphics[width = \textwidth, trim=0cm 0cm 0cm 0cm, clip]{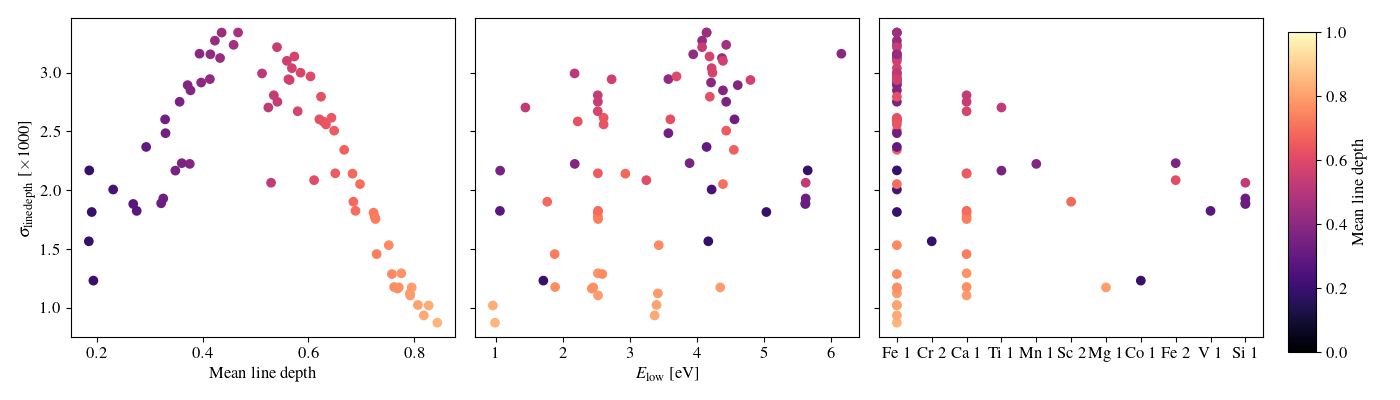}

    \caption{RMS of the temporal evolution of line depth at disc centre, for reconstructed lines in the wavelength region $5500-5600$~\AA~ and $6100-6200$~\AA~. From left to right, $\sigma_{RV, \mathrm{linedepth}}$ is plotted, and colour coded, against mean line depth, lower excitation potential and element + ionisation stage.}
    \label{fig:ld_jitter}
\end{figure*}

Lastly, we investigated the temporal variability of line depth in Figure \ref{fig:ld_jitter}. We find no monotonic relation with the atomic parameters, but do find an interesting parabolic-like relationship with mean line depth. Since the deep lines in our sample are likely to be saturated, we expect most of the variability to happen in the line wings instead of the line core. This is reflected by the large rms in equivalent width. On the other hand, the decreasing variability for shallow lines is due to Figure \ref{fig:ld_jitter} showing the variability in line depth as an absolute value. If instead we normalise $\sigma_\mathrm{linedepth}$ by the mean line depth itself, we recover a significant monotonic relationship where the most shallow lines show the largest relative variability, see Figure \ref{fig:tempvar_REW} in the Appendix.

\subsection{Centre-to-limb variation}\label{sec:CLV}
\begin{figure*}
    \centering
    \includegraphics[width = \textwidth, trim=0cm 0cm 0cm 0cm, clip]{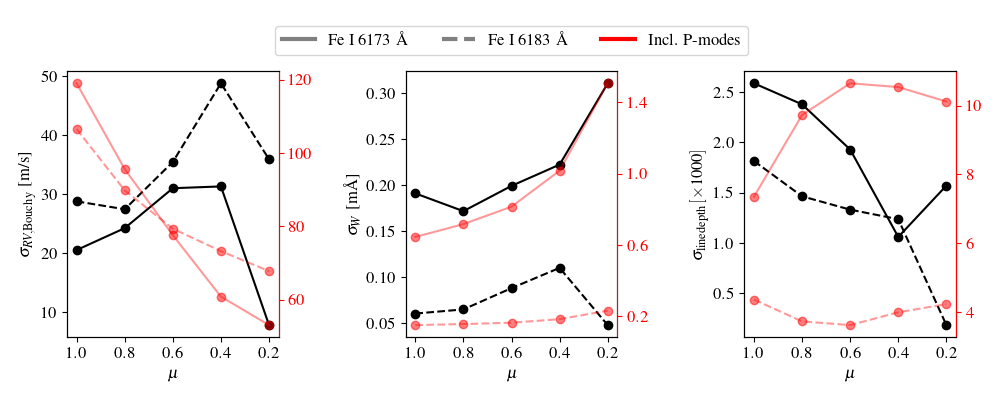}
    \caption{Centre-to-limb variation of the rms in $RV_\mathrm{Bouchy}$, equivalent width and line depth, for a strong (solid) and weak (dashed) \ion{Fe}{I} line. The values obtained from the reconstructed spectra (only granulation) and the original spectra (including p-modes) are shown in black and red, respectively.}
    \label{fig:CLV}
\end{figure*}

In Figure \ref{fig:CLV} we show the change in rms of $RV_\mathrm{Bouchy}$, equivalent width and line depth, from disc centre to the stellar limb for a representative strong line \ion{Fe}{I} 6173 \AA~and weak line \ion{Fe}{I} 6183 \AA. The results from the reconstructed spectra are shown in black, while those of the original spectra, still containing p-modes, are shown in red. It is clear from Figure~\ref{fig:CLV} that the presence of p-modes strongly alters the strength and limb-angle dependence of $\sigma_{RV_\mathrm{Bouchy}}$. The limb dependence of $\sigma_{RV_\mathrm{Bouchy}}$ of the reconstructed \ion{Fe}{I} 6173 \AA~line is similar to the one computed by \cite{frame2025}, with the variability being strongest at intermediate limb angles of $\mu\approx0.4-0.6$. In the solar photosphere, horizontal flows have larger velocities than vertical flows. In addition, as discussed in Section \ref{sec:results_RV}, the velocity field fluctuations decrease with height above the photosphere. Moreover, spectral lines at the stellar limb form at higher altitudes compared to lines forming at disc centre. Combining this information allows us to explain the CLV behaviour of $\sigma_{RV_\mathrm{Bouchy}}$ in Figure \ref{fig:CLV}. From disc centre to the limb, the spectral lines become increasingly sensitive to the horizontal flows, leading to an increase in $\sigma_{RV_\mathrm{Bouchy}}$. Simultaneously, spectral lines are formed at increasingly higher altitudes, leading to reduction in $\sigma_{RV_\mathrm{Bouchy}}$ at the smallest limb angles.

In the presence of p-modes, strong lines have larger RV variability at disc centre compared to weak lines, whereas the opposite is true near the limb. These results corroborate earlier findings by \cite{Dravins2023}, who studied the RV rms of idealised \ion{Fe}{I} lines at varying limb angles synthesised from a 3D HD model. Importantly, their synthetic spectra still includes the effect of p-modes. Our results are qualitatively the same as those shown in their Figure 10 \citep{Dravins2023}, although the absolute values of their RV rms ranges from $\sigma_{RV}=100-250~\ms$, from limb to disc-centre. The origin of this difference is not immediately clear. The 3D \texttt{CO5BOLD} models of \cite{Dravins2023} are set on a smaller box of size $5.6\times5.6\times2.25~\mathrm{Mm}$, compared to our \texttt{MURaM} model with size $9\times9\times5~\mathrm{Mm}$. As such, the p-modes excited in each respective model atmosphere will be different. In addition, since the model used in this work has a larger surface area, hence more visible granulation, there will be a difference in observed variability.

Looking at the limb dependence of variability in equivalent width and line depth in Figure \ref{fig:CLV}, we observe distinct difference between the shown spectral lines. The strong \ion{Fe}{I} 6173 \AA~exhibits an increase of variability in equivalent width  with decreasing limb angles. On the other hand, the weaker \ion{Fe}{I} 6183 \AA~line has a lower $\sigma_W$ than the strong line across all limb angles, and showing a decrease at $\mu=0.2$. Including p-modes does not significantly alter the shape of the limb-dependence, but does increase the size of the equivalent width variability.

\subsection{Correlations between RV and line diagnostics}\label{sec:corr_EW_RV}
In the previous sections, we discussed the size of the temporal variability in RV, equivalent width and line depth. In this section we investigate the instantaneous relation between these three line properties. Previous work by \cite{cegla2019} and \cite{Frame26} found that the granulation-induced RV is strongly correlated with the line equivalent width and line depth. Here we expand their analyses to investigate how these correlations behave when looking at a larger line sample, including different elements and ionisation stages.

\begin{figure*}
    \centering
    \includegraphics[width = \textwidth]{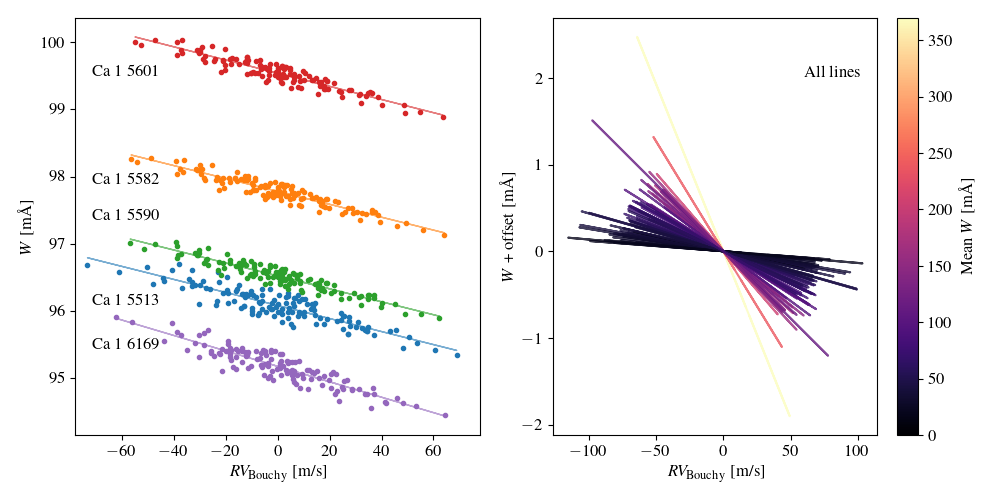}
    \caption{Equivalent width as a function of $RV_\mathrm{Bouchy}$ for each point in the reconstructed timeseries at disc centre. A subset of five lines are shown in the left panel, with their best fitting linear regression model shown as a solid line. In the right panel, the best fitting linear regression model for all lines are shown, offset by the intercept for ease of comparison. All linear regressions in the right panel are colour coded by their mean equivalent width.}
    \label{fig:corr_ew_rv}
\end{figure*}

\begin{figure*}
    \centering
    \includegraphics[width = \textwidth]{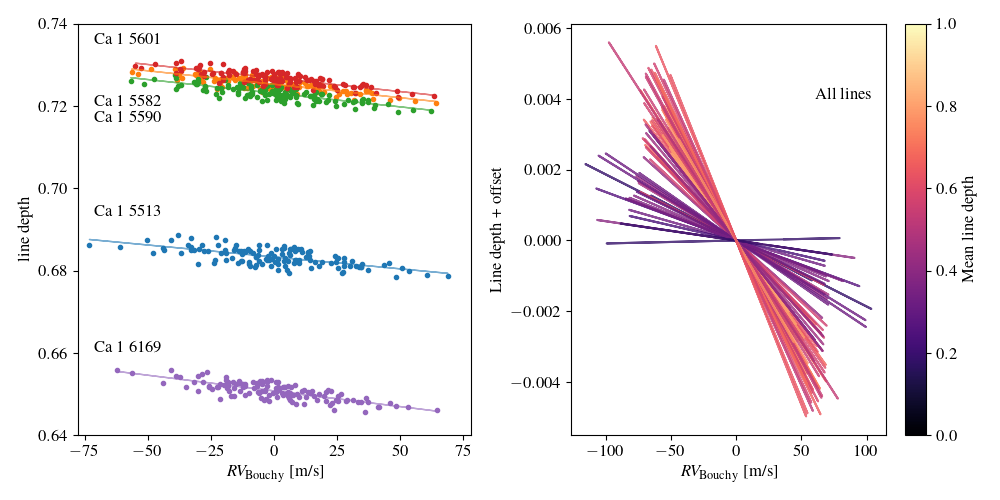}
    \caption{Spectral line depth as a function of $RV_\mathrm{Bouchy}$ for each point in the reconstructed timeseries at disc centre. A subset of five lines are shown in the left panel, with their best fitting linear regression model shown as a solid line. In the right panel, the best fitting linear regression model for all lines are shown, offset vertically by the intercept for ease of comparison. All linear regressions in the right panel are colour coded by their mean line depth.}
    \label{fig:corr_ld_rv}
\end{figure*}

In Figure \ref{fig:corr_ew_rv}, the line equivalent width is plotted as a function of $RV_\mathrm{Bouchy}$ for every point in the reconstructed timeseries at disc centre. Remember that $RV_\mathrm{Bouchy}$ is centred around the mean convective blueshift. This means that a value of zero is equal to the mean convective blueshift and not to the line rest wavelength. In the left panel a subset of five spectral lines are visualised. These are chosen for visualisation purposes, as they have similar line depths and equivalent widths. In addition, we performed linear regression analysis on the results for all spectral lines, shown as solid lines in Figure~\ref{fig:corr_ew_rv}, and computed their respective Spearman correlation coefficients. To facilitate a visual comparison of the correlation strengths between all spectral lines, we offset the linear regressions by subtracting their intercept in the right panel, such that they are centred around zero on the y-axis. Note that the units of the y-axis are the same for both panels such that the right panel shows the actual amplitude of the variability. A similar analysis is performed between line depth and $RV_\mathrm{Bouchy}$, as shown in Figure \ref{fig:corr_ld_rv}.

All lines show strong monotonic relations between equivalent width and $RV_\mathrm{Bouchy}$, with Spearman coefficients of $r_\mathrm{spearmann}<-0.6$. In general, the Spearman correlation decreases towards $r_\mathrm{spearmann}\rightarrow-1$ with increasing line strength. The steepness of the relation between equivalent width and RV also increases with increasing line strength. Besides, Figure \ref{fig:corr_ew_rv} reinforces the findings from Section \ref{sec:results_ind_lines} that weak lines have larger variability in RV, but smaller in equivalent width, than strong lines.

Regarding line depth, only the strongest spectral lines show significant monotonic relations with $RV_\mathrm{Bouchy}$; lines with depth $>0.75$ have $r_\mathrm{spearmann}<-0.75$. Weak to intermediate lines with line depth $<0.5$ show a wide range of correlation coefficients $-0.6<r_\mathrm{spearmann}<0.05$. This behaviour is not immediate evident from the regressions shown in the right panel of Figure \ref{fig:corr_ld_rv}, as the scatter around the regressions are not shown.

At the stellar limb ($\mu=0.4$), the situation stays similar for the relations between equivalent width and $RV_\mathrm{Bouchy}$. However, now all but three lines have strong monotonic relations between line depth and $RV_\mathrm{Bouchy}$, with Spearman coefficients $r_\mathrm{spearmann}<-0.6$. The three outliers have $-0.2<r_\mathrm{spearmann}<0.05$, and correspond to weak lines of \ion{V}{I}, \ion{Ti}{I} and \ion{Co}{I}.

\section{Does granulation affect spectral lines coherently?}\label{sec:results_coherence}

Coherent behaviour is important in the context of deriving RVs from real spectroscopic observations where most spectral lines will be blended with other lines. Specifically the strongest lines, which as shown in the previous sections have the most significant correlations between line shape and RV, are the most affected by line blending. If all lines, regardless of line strength, behave coherently, then line blending will only affect the absolute value of RV or equivalent width while keeping the temporal behaviour induced by granulation intact. For example, if at one instance of time granulation induces a blueshift in a spectral line, a coherent behaving line blend can only change the strength of the blueshift but never change it to a redshift.  

In addition to observational implications, the coherency between lines also allows us to expand the parametrisation method to create disc-integrated spectra. Since our 3D box-in-a-star models cover approximately $0.004\%$ of the solar surface, tiling a stellar disk with unique synthetic spectra at each spatial location would require up to $25\,000$ different snapshots. To circumvent this issue, \cite{Frame26} utilised their parametrisation method to develop \texttt{DISCO}, a publicly available code to create disc-integrated spectra of individual \ion{Fe}{I} lines. The method of \texttt{DISCO} is based on Eq. \ref{eq:reconstruction}, where filling factors are randomly sampled from their distribution to efficiently create new spectra. Each generated spectral line is unique but statistically consistent with the original spectral timeseries. Using this, \texttt{DISCO} creates a stellar disk with spatial resolution equal to that of our 3D models, having unique spectra at every grid point. As such, \texttt{DISCO} allows to study the temporal variability due to granulation on disc-integrated spectra, as compared to spatially resolved spectra used in this work. 

Due to the random nature of selecting filling factors, \texttt{DISCO} cannot create disc-integrated spectra for multiple lines that are parametrised separately. The reason is that each line has their own independent filling factor distributions. The resulting disc-integrated spectra of different spectral lines would therefore not represent the same underlying state of the stellar surface. However, if lines vary in phase, then we could expand the parametrisation method to classify multiple lines simultaneously; this would result in shared filling factor distributions between spectral lines. The advantage of such a multi-line method is that it would allow us to create disc-integrated spectra of larger spectral regions containing multiple lines. For this multi-line method to work, it is necessary that the spectral lines shifts, as caused by granulation, vary coherently in time.

\begin{figure}
    \centering
    \includegraphics[width = \columnwidth]{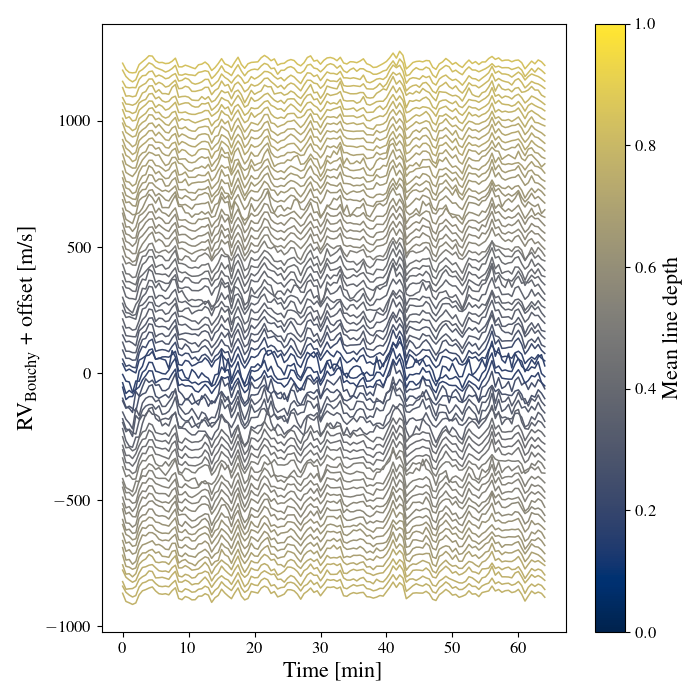}
    \caption[]{Temporal evolution of the RV computed for each spectral line plus an arbitrary offset of $\pm30~\ms$ to visualise the coherency between lines. The lines are sorted and colour coded by mean line depth.}
    \label{fig:coherence}
\end{figure}

In Figure \ref{fig:coherence} we have plotted the RV timeseries of every spectral line, offset by each other by an arbitrary value. We observed that the line diagnostics ($RV$, $W$, line depth) of most of the reconstructed spectral lines vary coherently in time, albeit with different amplitudes. Specifically, the majority of deep lines show strong coherent behaviour, while the shallowest lines behave the most different. This Figure corroborates the finding of \cite{Dravins2023} that most photospheric \ion{Fe}{I} and \ion{Fe}{II} lines of different strength vary in phase, with only small differences between spectral regions. 
 
It is then possible to select a coherent subsample to explore the feasibility of the multi-line parametrisation method. The line subsample was selected based on the 1-Wasserstein distance between the filling factor distributions of all lines. The 1-Wasserstein distance is representative of the cost of an optimal transport plan to convert one distribution into the other\footnote{\url{https://docs.scipy.org/doc/scipy/reference/generated/scipy.stats.wasserstein_distance.html}}. A small value of the 1-Wasserstein distance can be interpreted as two distributions being similar. As such, if two spectral lines have a similar filling factor distribution, it may be interpreted as the lines `seeing' the same surface granulation, i.e. they form at similar heights in the atmosphere.
%Another way to visualise this similarity is by plotting the line shift and equivalent width as a function of continuum intensity for each pixel in a 3D model atmosphere, see for example Figures 7 and 8 in \cite{Pereira09a}. 
In this work, we computed the 1-Wasserstein distance between the filling factor distributions of all lines within each spectral chunk, resulting in two 2D symmetric matrices, one for each chunk, where the diagonal is equal to zero. This is visualised in Figure \ref{fig:Wasserstein}. There is no clear cut-off value of the 1-Wasserstein distance that would represent similarity of two distributions. Therefore, we normalise the 1-Wasserstein values within each chunk to the largest value, to obtain a scale ranging between 0, most similar, and 1, most different. Next, we defined a subsample of `similar behaving' lines, within each spectral region, that have a normalised 1-Wasserstein distance smaller than 0.4. The resulting selection contains 48 spectral lines of only \ion{Fe}{I} and \ion{Ca}{I}, of which 28 are located in the chunk $5500-5600$ \AA~and 20 in the chunk $6100-6200$ \AA.

We subsequently re-parametrised these two subsamples chunk-by-chunk, as described in the last paragraph of Section \ref{sec:class}. As a result, in each chunk all the lines from the subsample share the same filling factors. In addition, rather than a single template line profile for each component, we then have template spectra that encompasses all the spectral lines of the subsample in each chunk. Using the new templates and filling factor distributions, we computed the reconstructed spectral timeseries using Eq. \ref{eq:reconstruction}. To validate that the new parametrisation is consistent with the previous line-by-line analysis, we compared the variability at disc centre in $RV_\mathrm{Bouchy}$, $RV_\mathrm{CCF}$, $RV_\mathrm{cog}$, line depth and $W$ of the spectral lines that are present in both reconstructed spectra. Ideally the difference in rms is as close to zero as possible. This would mean that both methods capture the same underlying line profile variability due to granulation. We find that the largest difference in rms, across all 48 spectral lines, is $0.85~\ms$, $0.88~\ms$, $1.02~\ms$, $0.00017$ and $0.01~$m\AA, for $RV_\mathrm{Bouchy}$, $RV_\mathrm{CCF}$, $RV_\mathrm{cog}$, line depth and $W$, respectively. For all variables the rms is larger in the line-by-line method as compared to the chunk-by-chunk method. These small differences in variability, relative to the values reported in Figures \ref{fig:RV_jitter} to \ref{fig:ld_jitter}, suggest that the chunk-by-chunk parametrisation can produce similar results as the line-by-line parametrisation, at least for the subsamples considered here. The advantage of the chunk-by-chunk method is that its filling factor distributions (of the three components) are equal for all the selected spectral lines. This in turn opens the door to creating disc-integrated spectra using \texttt{DISCO}, at least for the current selection of 48 \ion{Fe}{I} and \ion{Ca}{I} lines. In future work, when spectra for more spectral regions are available, it will be possible to extend these subsamples to include more lines.

\section{Discussion}\label{sec:discussion}

\subsection{Application to radial velocity observations}

Based on the results of \cite{cegla2019} and \cite{Frame26}, and those described in Section \ref{sec:corr_EW_RV}, we can conclude that the equivalent width of a line is linearly anti-correlated with its granulation-induced RV. On the other hand, the Doppler shift induced by a planet will not change the equivalent width of spectral lines. Hence, it may be feasible to perform signal analysis on both measured equivalent widths and RV's, to remove the common granulation signal, such as explored in \cite{Salzer25}. In order for such analysis to work, we will need to measure the equivalent width to a higher precision than its predicted variability. We can estimate the error on the measured equivalent width with the prescriptions of \cite{Norris01} or \cite{Cayrel1988,Cayrel04}. 

The Norris error is defined as:
\begin{equation}
    \sigma_{W,\mathrm{Norris}} = \lambda_0\cdot\sqrt{n_\mathrm{pix}}/(R\cdot\mathrm{SNR})~~,
\end{equation}
with $\lambda_0$ the centre wavelength of the line, $n_\mathrm{pix}$ the number of pixels that cover the line, $R$ the spectral resolution and $\mathrm{SNR}$ the signal-to-noise. The Cayrel error is defined as:
\begin{equation}
    \sigma_{W,\mathrm{Cayrel}} = \frac{1.5}{\mathrm{SNR}}\sqrt(FWHM\cdot\delta_x)~~,
\end{equation}
with $FWHM$ the full width at half maximum of the line and $\delta_x$ the pixel sampling.

Values typical of ultra-high resolution observations taken by the \texttt{ESPRESSO} spectrograph for bright targets are: $R=190\,000$, $\mathrm{SNR}=400$, $\lambda_0=6173~$\AA~ and $n_\mathrm{pix}\approx12$. Plugging in these numbers, we find equivalent width errors of $\sigma_{W,\mathrm{Norris}} =0.285~$m\AA~and $\sigma_{W,\mathrm{Cayrel}} =0.214~$m\AA. These errors are of similar size as the variability in equivalent width at disc centre shown in Figure \ref{fig:EW_jitter}. For example, $\sigma_W=0.15$~m\AA~and $\sigma_W=0.19$~m\AA~for the \ion{Fe}{I} 6151 and 6173 \AA~lines, respectively. To estimate the equivalent width rms of disc-integrated spectra, we use the \texttt{DISCO} code \citep{Frame26}, to generate disc-integrated spectral lines for \ion{Fe}{I} 6151 and 6173 \AA. The resulting variability is $\sigma_W=0.001$~m\AA~and $\sigma_W=0.0013$~m\AA~for the two lines, respectively. These values are a factor $\sim150$ smaller than the disc centre results, which is in-line with previous estimations by \cite{Dravins2023}.

From these estimates, it seems unfeasible to observe the variability in equivalent width from disc-integrated spectra using current high-resolution spectrographs. However, combining equivalent widths of several hundred strong spectral lines, similar to the determination of RV's in the line-by-line technique \citep{Dumusque_2018}, could increase the precision. This could be possible by defining a selection of spectral lines that behave coherently in time, such that their equivalent widths and RV measurements can be co-added. On the other hand, spatially resolved solar observations using \texttt{PoET} \citep{Santos25} should be able to resolve the equivalent width variability due to granulation, at disc centre and the solar limb. 

%\cis{Not the biggest fan of the Cayrel error as it requires a measure of the FWHM. So may remove that one altogether. For the Norris error, I am not too sure about the number of pixels that covers a typical line. Now I hand-waved a number based on the resolution, mask of 0.2\AA~ and 2 pixel/element sampling.}

\subsection{Dependence on RV method}\label{sec:RV_method_diff}

RV is often treated as a physical concept, describing the velocity of a system along the line-of-sight. However, the quantity itself, as derived from observed spectra, is method dependent. In most studies, RV's are computed by fitting a Gaussian to the CCF and taking its centre value, where the CCF is derived from a masked observed spectrum and a reference template (spectrum). In addition, the CCF is computed from many hundreds of spectral lines, curated to exclude tellurics and strong blends, returning a single value for the RV. More recently, \cite{Dumusque_2018} proposed a line-by-line approach, where an RV is computed for each line separately using a template matching method \citep{Bouchy01}. All the individual RV's are subsequently combined in a weighted average, with the respective RV error of each line as weight, to result in a single final RV. A less common metric for the RV is the first moment of the line profile intensity, or line centre-of-gravity ($RV_\mathrm{cog}$), which represents the intensity-weighted velocity of a spectral line with respect to its rest wavelength. Following its definition in Eq. \ref{eq:lambda_cog}, it requires both edges of the spectral region, that is being integrated over, to reach continuum. Otherwise, the resulting RV will be biased to either the blue or the red. 

First, we investigated the impact of the integration mask used to compute $RV_\mathrm{cog}$, with respect to the accuracy of $RV_\mathrm{cog}$. Our test setup consists of a Gaussian, mimicking the line depth and width of the solar Fe I 6173 \AA~line at a resolution of $R=2~000~000$, blueshifted in steps of $0.1~\ms$ from $-1$ to $-3~\mathrm{km~s^{-1}}$. The red edge of the Gaussian reaches far in the continuum ($=1$) while the blue edge is cut off close to the line wing. This ensures that with increasing blueshift, the blue edge will reach increasingly lower flux levels below the continuum. At each step we calculated $RV_\mathrm{cog}$ and computed the error relative to the true blueshift. We find that the error is larger than $\gtrapprox10~\cms$ if the blue edge has a flux value smaller than $\lessapprox0.99994$, compared to the continuum value of $1$ at the red edge. The error increases to $\gtrapprox1$ and $10~\ms$ with the blue edge at a flux level of $\lessapprox0.9994$ and $0.994$, respectively. In practice it will be difficult to define line masks that have equal continuum values to a precision of $5\times10^{-3}$ between both edges. We conclude that using $RV_\mathrm{cog}$ for parametrisation purposes as done in this work, where a precision of $~100~\ms$ is sufficient, is fine. However, utilising $RV_\mathrm{cog}$ to characterise granulation induced RV's, and ultimately detect planetary signal at a sub meter-per-second precision, seems unfeasible.

\begin{figure}
    \centering
    \includegraphics[width = \columnwidth]{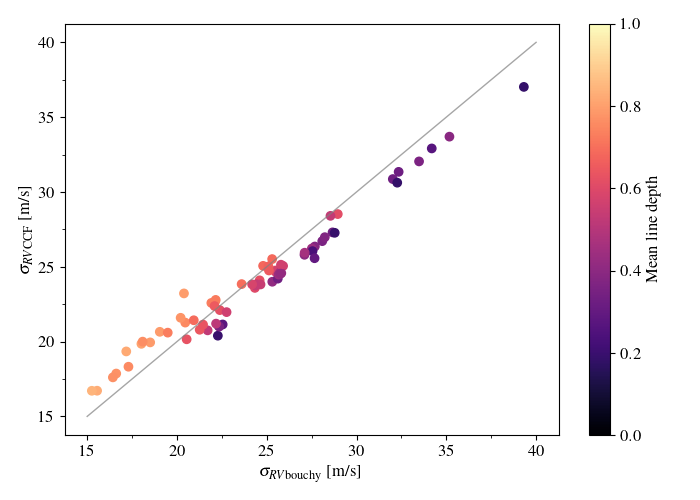}    %\includegraphics[width = \columnwidth]{plots/comp_RV_M1_Bouchy.png}

    \caption{Comparison of the temporal RV rms computed using $RV_\mathrm{CCF}$ and the template matching technique from \citet{Bouchy01}, colour-coded by mean line depth.}
    \label{fig:RVrms_comp}
\end{figure}

Next, we computed the difference in derived RV between the CCF and template matching methods at disc centre, for all spectral lines ($N_\mathrm{lines}=72$) and snapshots in time ($N_\mathrm{timesteps}=129$). The difference in RV rms over the full timeseries at disc centre is shown in Figure \ref{fig:RVrms_comp}, and is at most $\approx2.5~\ms$.  The absolute difference between the instantaneous $RV_\mathrm{CCF}$ and $RV_\mathrm{Bouchy}$ measurements is at most $\lessapprox\pm7.5~\ms$, with the majority of points having a difference less than $\lessapprox\pm2.5~\ms$. The largest absolute differences occur when the spectral lines reach their largest line shifts ( $RV_\mathrm{Bouchy}\approx100~\ms$, see the x-axis of Figure \ref{fig:corr_ew_rv}). However, the relative difference as measured against $RV_\mathrm{Bouchy}$ blows up when $RV_\mathrm{Bouchy}$ becomes smaller than $5~\ms$, ranging from $\pm20\%$ up to $\pm100\%$. Comparing to a realistic example, \cite{Dumusque_2018} computed RV's for individual spectral lines using the template matching method down to a precision of at best $\sim10~\ms$. Taking $RV_\mathrm{Bouchy}=10~\ms$, the associated relative difference between a CCF based method and template matching method is at most $<\pm 15\%$. 

At the stellar limb ($\mu=0.4$), the largest difference in RV is $\lessapprox\pm10~\ms$ for the strongest lines, with most other lines having a difference in RV less than $\lessapprox2.5~\ms$. Similar to disc centre, the relative difference increases substantially for small values of $RV_\mathrm{Bouchy}<5~\ms$, from $20\%$ up to $100\%$. Regarding the RV rms at the stellar limb, the difference is at most $<4~\ms$.   %\cis{Not sure how much this changes when going to disc-integrated and mimicking longer exposures. Both will result in smaller RV's on the order of cm/s, the question is, will the relative difference between methods stay? Could potentially check with the DISCO code?}

We can draw three conclusions regarding the use of different RV methods, based on the results shown in this Section and Section \ref{sec:results_ind_lines}. First, the line centre-of-gravity is strongly dependent on the line masks used to perform its summation, and hence unfeasible in practice to obtain RV's to sub ten meter-per-second accuracy. Secondly, we find that the difference in RV's measured between the CCF and template matching method can become substantial when the line shifts become small. If future instruments push the precision of individual line RV measurements below $10~\ms$, a more comprehensive comparison of methods is needed. Thirdly, a line-by-line approach is favoured over a full-spectrum CCF approach when the goal is to study or minimise RV variability from granulation; this is because different lines will experience RV shifts with different amplitudes over time. Since the CCF is computed by calculating the integral of the product between two spectra at fixed velocity steps, it naturally muddles RV information from different spectral lines, removing any coherency that previously existed. Contrary to the CCF, the line-by-line approach does not suffer from this issue.

\section{Summary and conclusions}
In this work we have utilised the parametrisation technique from \cite{frame2025} to investigate how granulation affects the shape and shifts of 72 spectral lines in the wavelength regions $5500-5600$ \AA~and $6100-6200$ \AA, in a non-magnetic 3D solar model atmosphere. This is the largest and most diverse sample of spectral lines studied in this context so far, including 11 different elements of various ionisation stages and line strengths. By using our parametrisation technique, we can remove p-modes from synthetic spectra, fully isolating the granulation-induced spectral effects.

First, we find that both at disc-centre and the stellar limb ($\mu=0.4$) the temporal variability in granulation-induced RV is strongly correlated with line depth and lower excitation potential. The variability in equivalent width is equally strongly correlated with line depth, but not with lower excitation potential. Weaker lines exhibit greater RV variability, but smaller variability in equivalent width, than strong lines. The amplitude of the variability increases from disc centre to the limb; for example, for weak lines: $\sigma_{RV}\sim20-40~\ms$ and $\sigma_{W}\sim0.03-0.18~$m\AA~at disc centre, and $\sigma_{RV}\sim30-50~\ms$ and $\sigma_{W}\sim0.01-0.30~$m\AA~at a limb angle of $\mu=0.4$. On the other hand, for strong lines: $\sigma_{RV}\sim15-25~\ms$ and $\sigma_{W}\sim0.14-1.0~$m\AA~at disc centre, and $\sigma_{RV}\sim11-40~\ms$ and $\sigma_{W}\sim0.12-1.05~$m\AA~at a limb angle of $\mu=0.4$. Moreover, we conclude that the presence of p-modes strongly alters the centre-to-limb variation of the RV and equivalent width rms. Without p-modes, strong lines exhibit larger variability in both RV and equivalent width across all limb angles, and the RV rms is largest at the stellar limb ($\mu\approx0.4-0.6$) for all lines. On the other hand, when p-modes are present, the RV rms monotonically decreases for all lines towards the stellar limb, with the strongest lines having larger RV variability at disc centre, but weaker at the stellar limb, than weak lines. 

Secondly, there are clear linear correlations between the instantaneous line equivalent width, depth and RV, as caused by stellar surface granulation. The strength of these correlations depend greatly on line depth and limb angle, with stronger lines exhibiting the strongest correlations at all limb angles. \cite{cegla2019} and \cite{Frame26} previously used such relations, for four individual \ion{Fe}{I} lines, to remove granulation noise, finding reductions in RV rms up to 60\% in ideal noise-free circumstances. Here we show that the strongest spectral lines are the most promising candidates for this kind of noise reduction. Especially, since observing variability in equivalent width for individual spectral lines may be too challenging for all but the strongest lines. Another potential avenue for planet detection, is to combine information of the equivalent width and RV variability in frequency space. Since only the RV is affected by the planet-induced Doppler shift it may be possible to remove the common granulation signal and extract the underlying planetary radial-velocity signal.

Thirdly, we found that the granulation-induced RV and equivalent width of most spectral lines in our sample vary coherently in time. This has the import conclusion that line blends should not distort the granulation signal, but only change the absolute derived values of RV and equivalent width. In addition, the coherence across numerous spectral lines allowed us to identify a subsample of  \ion{Fe}{I} and \ion{Ca}{I} lines which filling factor distributions are statistically similar. Subsequently parametrising all the lines in this subsample simultaneously, opens the door to creating disc-integrated spectra for multiple spectral lines using the \texttt{DISCO} code.  

In subsequent work, we will expand our sampled wavelength regions and stellar limb angles, and compute spectra from MHD model atmospheres with a small-scale dynamo and larger uniform magnetic field strengths upwards of $B>200~\mathrm{G}$. By studying the effect of small- and large-scale magnetic fields, we will be able to study the impact of magnetic bright points and faculae on spectral line formation, and the resulting RV variability. This will bring theoretical work closer to real observations, help us find mitigation strategies to remove the stellar variability from observations, and ultimately, enable the detection and characterisation of earth-analogues.

\section{Acknowledgments}
We thank the anonymous referee for their constructive feedback which has helped to improve the manuscript. CL and HC acknowledge funding from a UKRI Future Leader Fellowship (grant numbers MR/S035214/1 and MR/Y011759/1). GF acknowledges a Warwick prize scholarship (PhD) made possible thanks to a generous philanthropic donation. C.A.W. would like to acknowledge support from the UK Science and Technology Facilities Council (STFC, grant number ST/X00094X/1). VW and AS acknowledge support from the European Research Council (ERC) under the European Union’s Horizon 2020 research and innovation program (grant No. 101118581—project REVEAL). This work has made use of the VALD database, operated at Uppsala University, the Institute of Astronomy RAS in Moscow, and the University of Vienna. Computing facilities were provided by the Scientific Computing Research
Technology Platform of the University of Warwick.

\section{Data Availability}

Datasets generated during this study are available upon reasonable request.



\bibliographystyle{mnras}
\bibliography{bib} % if your bibtex file is called example.bib

%\begin{thebibliography}{99}
%\end{thebibliography}

%%%%%%%%%%%%%%%%%%%%%%%%%%%%%%%%%%%%%%%%%%%%%%%%%%

\appendix 

\section{Atomic data}\label{App:AtomicData}

Table \ref{tab:atomic_data1} and \ref{tab:atomic_data2} list the atomic properties of all the spectral lines that were synthesised in this work. All data has been compiled from the VALD database, individual references are listed in the Table's caption.
\begin{table*}
    \centering
    \begin{tabular}{l c c c c c c c c}
    \hline
    Element + ionisation & $\lambda_\mathrm{air}$ (nm) & log($gf$) & Excitation potential (eV)&  $\gamma_{\text{rad}}$ & $\gamma_{\text{Stark}}$ & $\gamma_{\text{VdW}}$ & Sources$^{\dagger}$\\
    \hline
    \ion{Fe}{1}  & 549.752 & -2.849 & 1.011 & 7.16 & -6.22 & -7.82 & 1,2   \\
    \ion{Fe}{1}  & 550.146 & -3.047 & 0.958 & 7.19 & -6.22 & -7.82 & 1,3  \\
    \ion{Cr}{2}  & 550.209 & -1.96 & 4.168 & 8.37 & -6.49 & -7.86 & 4,5,6 \\ 
    \ion{Fe}{1}  & 550.307 & -1.417 & 4.386 & 8.3 & -4.45 & -7.51 & 1   \\
    \ion{Fe}{1}  & 550.678 & -2.797 & 0.99 & 7.17 & -6.22 & -7.82 & 1,2   \\
    \ion{Ca}{1}  & 551.298 & -0.464 & 2.932 & 8.447 & -4.053 & -7.316 & 7 \\
    \ion{Ti}{1}  & 551.453 & -0.5 & 1.443 & 8.15 & -6.07 & -7.71 & 8,9,6      \\
    \ion{Mn}{1}  & 551.677 & -1.847 & 2.178 & 8.62 & -6.0 & -7.74 & 10,11      \\
    \ion{Fe}{1}  & 551.957 & 0.34 & 6.15 & 8.3 & -5.15 & -7.51 & 1   \\
    \ion{Fe}{1}  & 552.245 & -1.55 & 4.209 & 8.01 & -5.62 & -7.53 & 1,2  \\
    \ion{Fe}{1}  & 552.398 & -1.346 & 4.558 & 8.31 & -5.01 & -7.31 & 1 \\
    \ion{Fe}{1}  & 552.554 & -1.084 & 4.23 & 8.47 & -5.4 & -7.53 & 1,12 \\
    \ion{Sc}{2}  & 552.679 & -0.01 & 1.768 & 8.35 & -6.57 & -7.81 & 13,14,15 \\
    \ion{Mg}{1}  & 552.84 & -0.498 & 4.346 & 8.72 & -4.46 & -6.979 & 16,17,18 \\
    \ion{Co}{1}  & 553.078 & -2.06 & 1.71 & 7.71 & -6.2 & -7.81 & 19,2 \\
    \ion{Fe}{1}  & 553.275 & -2.15 & 3.573 & 7.8 & -5.86 & -7.79 & 1,2 \\
    \ion{Fe}{2}  & 553.484 & -2.73 & 3.245 & 8.46 & -6.52 & -7.89 & 20,21 \\
    \ion{Fe}{1}  & 553.542 & -1.158 & 4.186 & 7.81 & -5.88 & -7.8 & 1,3 \\
    \ion{Fe}{1}  & 554.315 & -1.57 & 3.695 & 7.13 & -6.11 & -7.82 & 1,2 \\
    \ion{Fe}{1}  & 554.394 & -1.14 & 4.218 & 8.47 & -5.62 & -7.53 & 1,2  \\
    \ion{Fe}{1}  & 554.651 & -1.31 & 4.371 & 8.31 & -5.28 & -7.51 & 1,2  \\
    \ion{Fe}{1}  & 554.699 & -1.91 & 4.218 & 8.54 & -5.44 & -7.57 & 1,2  \\
    \ion{Fe}{1}  & 555.358 & -1.41 & 4.435 & 8.28 & -4.94 & -7.51 & 1,2  \\
    \ion{Fe}{1}  & 555.489 & -0.44 & 4.548 & 8.21 & -4.97 & -7.42 & 1,2  \\
    \ion{Fe}{1}  & 556.021 & -1.19 & 4.435 & 8.28 & -4.38 & -7.51 & 1,2  \\
    \ion{Fe}{1}  & 556.124 & -1.204 & 4.607 & 8.29 & -4.96 & -7.31 & 1  \\
    \ion{Fe}{1}  & 556.211 & -0.436 & 4.386 & 8.31 & -5.01 & -7.51 & 1 \\
    \ion{Fe}{1}  & 556.271 & -0.642 & 4.435 & 8.28 & -4.66 & -7.52 & 1 \\
    \ion{Fe}{1}  & 556.36 & -0.99 & 4.191 & 8.46 & -5.33 & -7.53 & 1,2 \\
    \ion{Fe}{1}  & 556.739 & -2.564 & 2.608 & 8.2 & -5.88 & -7.73 & 1,22 \\
    \ion{Fe}{1}  & 556.962 & -0.486 & 3.417 & 8.06 & -5.38 & -7.54 & 1,12 \\
    \ion{Fe}{1}  & 557.284 & -0.275 & 3.396 & 8.06 & -5.41 & -7.54 & 1,22 \\
    \ion{Fe}{1}  & 557.609 & -1.0 & 3.43 & 8.06 & -5.39 & -7.54 & 1,2 \\
    \ion{Ca}{1}  & 558.196 & -0.555 & 2.523 & 7.853 & -6.072 & -7.628 & 23 \\
    \ion{Fe}{1}  & 558.427 & -0.988 & 4.386 & 8.3 & -5.28 & -7.51 & 1 \\
    \ion{Fe}{1}  & 558.476 & -2.32 & 3.573 & 8.23 & -5.75 & -7.74 & 1,2 \\
    \ion{Fe}{1}  & 558.676 & -0.12 & 3.368 & 8.06 & -5.38 & -7.54 & 1,3 \\
    \ion{Fe}{1}  & 558.757 & -1.85 & 4.143 & 7.68 & -6.02 & -7.8 & 1,2 \\
    \ion{Ca}{1}  & 558.875 & 0.358 & 2.526 & 7.853 & -6.072 & -7.628 & 23 \\
    \ion{Ca}{1}  & 559.011 & -0.571 & 2.521 & 7.852 & -6.071 & -7.628 & 23 \\
    \ion{Ca}{1}  & 559.446 & 0.097 & 2.523 & 7.852 & -6.072 & -7.628 & 23 \\
    \ion{Ca}{1}  & 560.128 & -0.523 & 2.526 & 7.852 & -6.072 & -7.628 & 23 \\
    \hline
    \end{tabular}
    \caption{Atomic parameters for the lines selected for this work in the wavelength region $5500-5600$~\AA. $\gamma_{\text{rad}}$, $\gamma_{\text{Stark}}$ and $\gamma_{\text{VdW}}$ refer to the radiative, Stark and Van der Waals damping parameters. All data is sourced from the VALD database. \\[0.5ex]
    $^{\dagger}$ Sources: \textbf{(1)} \protect\cite{K14,Kurucz2018}; \textbf{(2)} \protect\cite{FMW}; \textbf{(3)} \protect\cite{BWL}; \textbf{(4)} \protect\cite{SNave}; \textbf{(5)} \protect\cite{LSNE}; \textbf{(6)} \protect\cite{K16,Kurucz2018}; \textbf{(7)} \protect\cite{S}; \textbf{(8)} \protect\cite{Sal12a}; \textbf{(9)} \protect\cite{LGWSC}; \textbf{(10)} \protect\cite{K07,Kurucz2018}; \textbf{(11)} \protect\cite{MFW}; \textbf{(12)} \protect\cite{BK}; \textbf{(13)} \protect\cite{NIST22}; \textbf{(14)} \protect\cite{LHSNWC}; \textbf{(15)} \protect\cite{K09,Kurucz2018}; \textbf{(16)} \protect\cite{NIST10}; \textbf{(17)} \protect\cite{LZ}; \textbf{(18)} \protect\cite{BPM}; \textbf{(19)} \protect\cite{K08,Kurucz2018}; \textbf{(20)} \protect\cite{K13,Kurucz2018}; \textbf{(21)} \protect\cite{BSScor}; \textbf{(22)} \protect\cite{BKK}; \textbf{(23)} \protect\cite{SR}.}
    \label{tab:atomic_data1}
\end{table*}

\begin{table*}
    \centering
    \begin{tabular}{l c c c c c c c}
    \hline
    Element + ionisation & $\lambda_\mathrm{air}$ (nm) & log($gf$) & Excitation potential (eV) & $\gamma_{\text{rad}}$ & $\gamma_{\text{Stark}}$ & $\gamma_{\text{VdW}}$ & Sources$^{\dagger}$ \\
    \hline
    \ion{Ca}{1}  & 610.272 & -0.793 & 1.879 & 7.86 & -5.32 & -7.189 & 1 \\
    \ion{Fe}{1}  & 610.732 & -1.577 & 4.076 & 7.91 & -6.03 & -7.79 & 2  \\
    \ion{V }{1}  & 611.953 & -0.36 & 1.064 & 7.66 & -6.07 & -7.77 & 3,4,5\\
    \ion{Ca}{1}  & 612.222 & -0.316 & 1.886 & 7.86 & -5.32 & -7.189 & 1 \\
    \ion{Si}{1}  & 612.502 & -1.465 & 5.614 & 7.91 & -5.56 & -7.13 & 6 \\
    \ion{Ti}{1}  & 612.622 & -1.206 & 1.067 & 6.92 & -6.1 & -7.77 & 7 \\
    \ion{Fe}{1}  & 612.791 & -1.399 & 4.143 & 8.0 & -6.02 & -7.79 & 2,8 \\
    \ion{Fe}{1}  & 613.661 & -1.4 & 2.453 & 7.96 & -6.08 & -7.77 & 2,9 \\
    \ion{Fe}{1}  & 613.769 & -1.403 & 2.588 & 8.08 & -6.29 & -7.75 & 2,9 \\
    \ion{Fe}{1}  & 614.173 & -1.459 & 3.602 & 8.08 & -5.41 & -7.54 & 2,10 \\
    \ion{Si}{1}  & 614.248 & -1.296 & 5.619 & 7.9 & -3.57 & -7.13 & 6 \\
    \ion{Si}{1}  & 614.502 & -1.311 & 5.616 & 7.92 & -3.63 & -7.13 & 6 \\
    \ion{Fe}{2}  & 614.925 & -2.72 & 3.889 & 8.5 & -6.53 & -7.88 & 11,12 \\
    \ion{Fe}{1}  & 615.162 & -3.299 & 2.176 & 8.29 & -6.16 & -7.79 & 2,9 \\
    \ion{Si}{1}  & 615.513 & -0.755 & 5.619 & 7.9 & -3.16 & -7.13 & 6 \\
    \ion{Fe}{1}  & 615.773 & -1.26 & 4.076 & 7.89 & -6.03 & -7.79 & 2,9 \\
    \ion{Ca}{1}  & 616.13 & -1.266 & 2.523 & 7.274 & -4.994 & -7.264 & 13 \\
    \ion{Ca}{1}  & 616.376 & -1.286 & 2.521 & 7.296 & -4.998 & -7.264 & 13 \\
    \ion{Fe}{1}  & 616.536 & -1.474 & 4.143 & 8.0 & -6.02 & -7.78 & 2,8 \\
    \ion{Ca}{1}  & 616.644 & -1.142 & 2.521 & 7.269 & -4.999 & -7.264 & 13 \\
    \ion{Ca}{1}  & 616.904 & -0.797 & 2.523 & 7.296 & -4.997 & -7.264 & 13 \\
    \ion{Ca}{1}  & 616.956 & -0.478 & 2.526 & 7.273 & -4.994 & -7.264 & 13 \\
    \ion{Fe}{1}  & 617.051 & -0.44 & 4.796 & 8.3 & -4.99 & -7.42 & 2,9 \\
    \ion{Fe}{1}  & 617.333 & -2.88 & 2.223 & 8.31 & -6.16 & -7.79 & 2,9 \\
    \ion{Fe}{1}  & 618.02 & -2.586 & 2.728 & 8.21 & -5.91 & -7.73 & 2,14 \\
    \ion{Fe}{1}  & 618.356 & -1.175 & 5.033 & 8.79 & -3.77 & -7.33 & 2 \\
    \ion{Fe}{1}  & 618.569 & -0.625 & 5.649 & 7.9 & -4.91 & -7.51 & 2 \\
    \ion{Fe}{1}  & 618.799 & -1.72 & 3.943 & 7.75 & -5.38 & -7.54 & 2,9 \\
    \ion{Fe}{1}  & 619.156 & -1.417 & 2.433 & 7.99 & -6.08 & -7.78 & 2,8 \\
    \ion{Fe}{1}  & 620.031 & -2.437 & 2.608 & 8.08 & -6.29 & -7.75 & 2,9 \\
    \hline
    \end{tabular}
    \caption{Atomic parameters for the lines selected for this work in the wavelength region $6100-6200$~\AA. $\gamma_{\text{rad}}$, $\gamma_{\text{Stark}}$ and $\gamma_{\text{VdW}}$ refer to the radiative, Stark and Van der Waals damping parameters. All data is sourced from the VALD database. \\[0.5ex]
    $^{\dagger}$ Sources: \textbf{(1)} \protect\cite{SN}; \textbf{(2)} \protect\cite{K14,Kurucz2018}; \textbf{(3)} \protect\cite{TPSa}; \textbf{(4)} \protect\cite{LWDFSC}; \textbf{(5)} \protect\cite{K09,Kurucz2018}; \textbf{(6)} \protect\cite{K07,Kurucz2018}; \textbf{(7)} \protect\cite{K16,Kurucz2018}; \textbf{(8)} \protect\cite{BWL}; \textbf{(9)} \protect\cite{FMW}; \textbf{(10)} \protect\cite{BKK}; \textbf{(11)} \protect\cite{K13,Kurucz2018}; \textbf{(12)} \protect\cite{BSScor}; \textbf{(13)} \protect\cite{SR}; \textbf{(14)} \protect\cite{BK}.}
    \label{tab:atomic_data2}
\end{table*}

\section{Bisector comparisons to the IAG Solar atlas}
\label{App:bisectors}

In Figure \ref{fig:bis_comp} we compared the bisectors of the \ion{Fe}{I} 6151 and 6173 \AA~lines to those by \cite{frame2025} and the IAG spatially-resolved solar atlas \citep{IAG}. Here we show the remaining bisector comparisons with IAG, computed from the mean of the disc-centre reconstructed timeseries of each line. Note that the bisectors have their mean value subtracted, so they are centred around zero. The IAG spectra is downgraded to match the resolution of the synthetic spectra ($R=500\,000$), and has removed telluric contamination from $\mathrm{H_2O}$ and $\mathrm{O_2}$ lines based on the \texttt{HITRAN2020} database. We find that the majority of bisectors match with the IAG atlas. Notable exceptions are the \ion{Fe}{I} 5503, 5555, 5567, 6171 \AA~, and the \ion{Si}{I} 6125, 6155 \AA~lines. We expect that these differences originate from the presence of line blends in the IAG data, which are absent from our synthetic spectra by construction. To investigate the impact of line blends, we would need to synthesise new spectra that includes a more comprehensive linelist. Even then, it is not a given that all the atomic data is sufficiently complete or accurate to reproduce all spectral lines. From the observational point of view, there could be residual tellurics that were not (fully) removed in the data processing.

\begin{figure*}
    \centering
    \includegraphics[width = \textwidth, trim=0cm 0cm 0cm 0cm, clip]{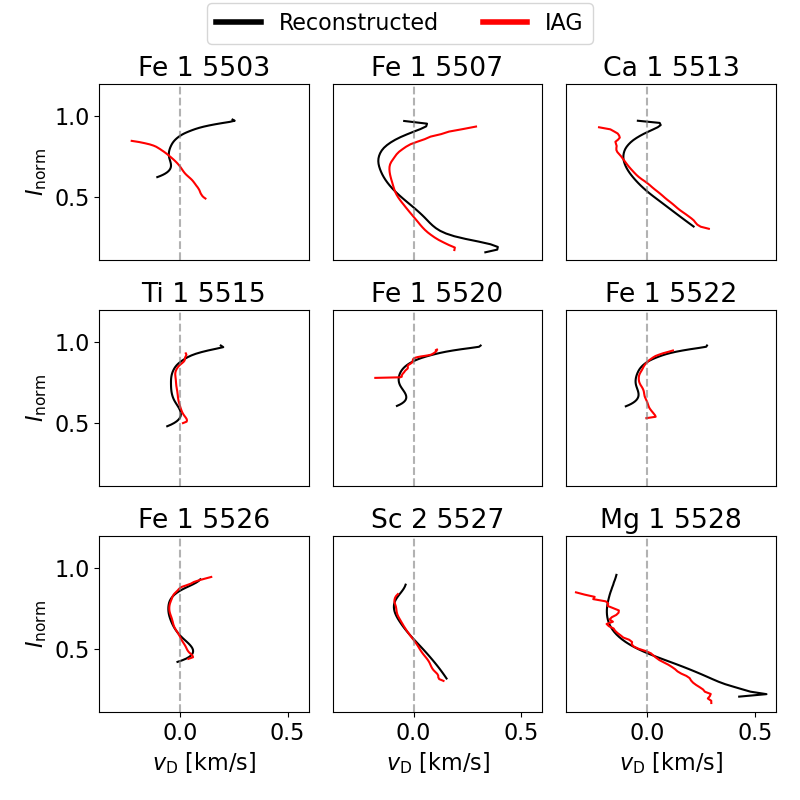}
    \caption{Line bisectors of the reconstructed spectra (black) as compared to the spatially resolved IAG atlas at disc centre (red). Each bisector has their mean subtracted to centre it on zero, and to facilitate a comparison of bisector shape.}
    \label{fig:ALLbis3}
\end{figure*}

\begin{figure*}
    \centering
    \includegraphics[width = \textwidth, trim=0cm 0cm 0cm 0cm, clip]{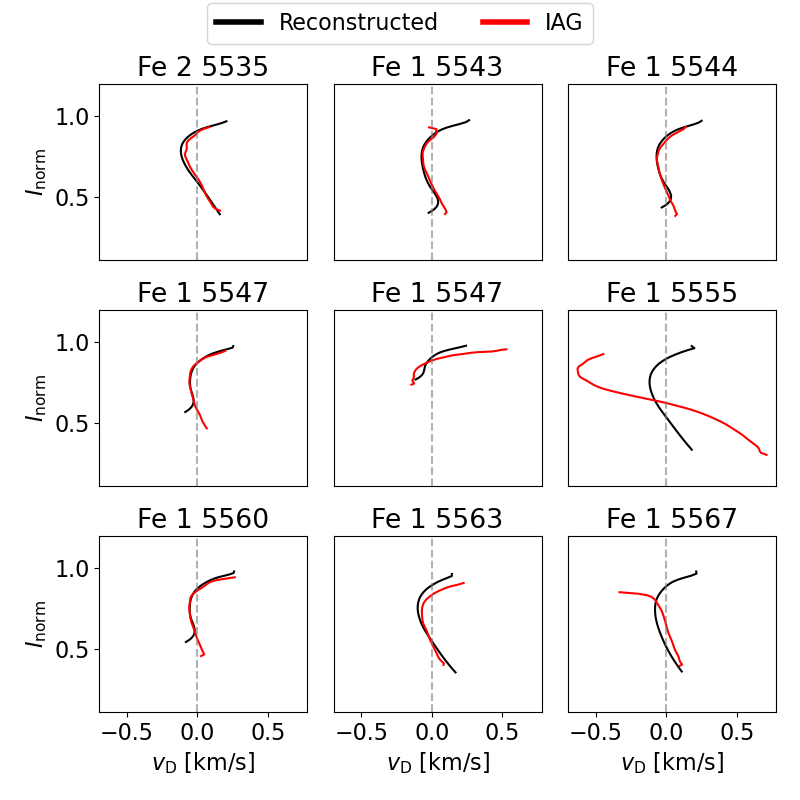}
    \caption{Line bisectors of the reconstructed spectra (black) as compared to the spatially resolved IAG atlas at disc centre (red). Each bisector has their mean subtracted to centre it on zero, and to facilitate a comparison of bisector shape.}
    \label{fig:ALLbis4}
\end{figure*}

\begin{figure*}
    \centering
    \includegraphics[width = \textwidth, trim=0cm 0cm 0cm 0cm, clip]{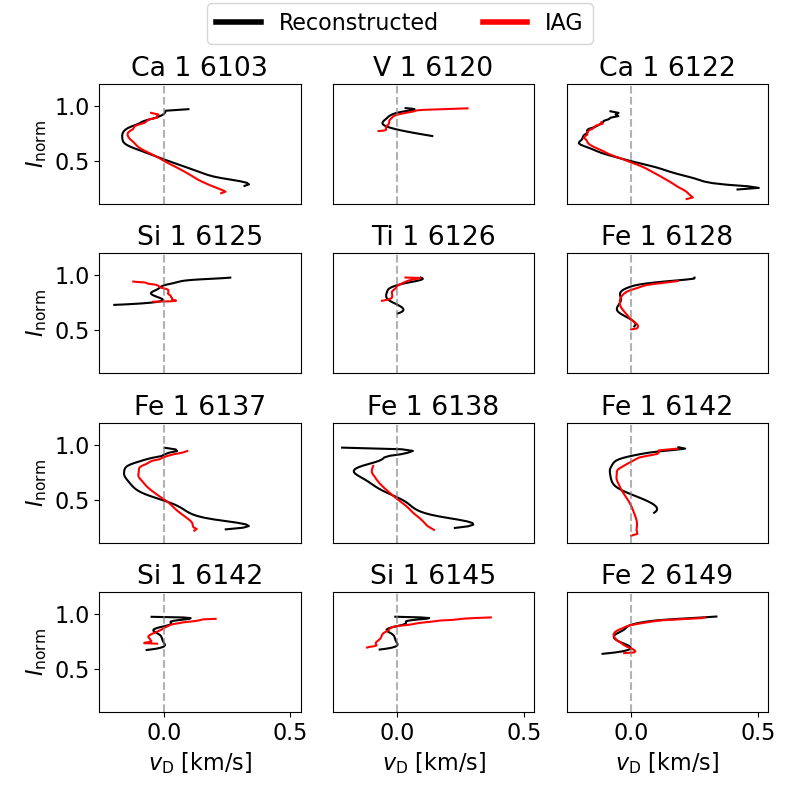}
    \caption{Line bisectors of the reconstructed spectra (black) as compared to the spatially resolved IAG atlas at disc centre (red). Each bisector has their mean subtracted to centre it on zero, and to facilitate a comparison of bisector shape.}
    \label{fig:ALLbis1}
\end{figure*}

\begin{figure*}
    \centering
    \includegraphics[width = \textwidth, trim=0cm 0cm 0cm 0cm, clip]{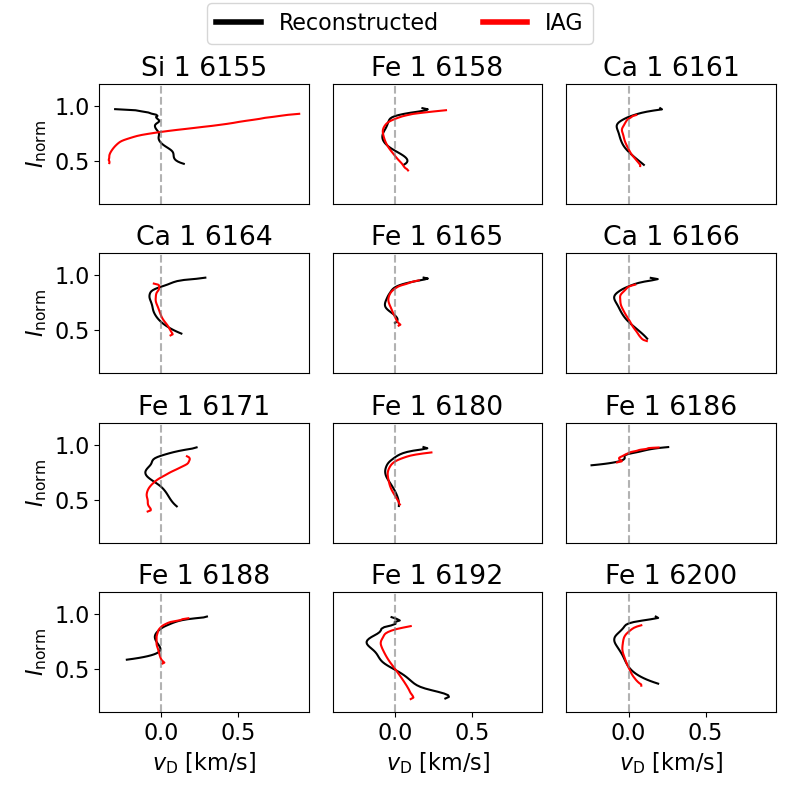}
    \caption{Line bisectors of the reconstructed spectra (black) as compared to the spatially resolved IAG atlas at disc centre (red). Each bisector has their mean subtracted to centre it on zero, and to facilitate a comparison of bisector shape.}
    \label{fig:ALLbis2}
\end{figure*}

\section{Hydrodynamic structure of the \texttt{MURaM} Solar simulation}

\begin{figure*}
    \centering
    \includegraphics[width = \textwidth, trim=0cm 0cm 0cm 0cm, clip]{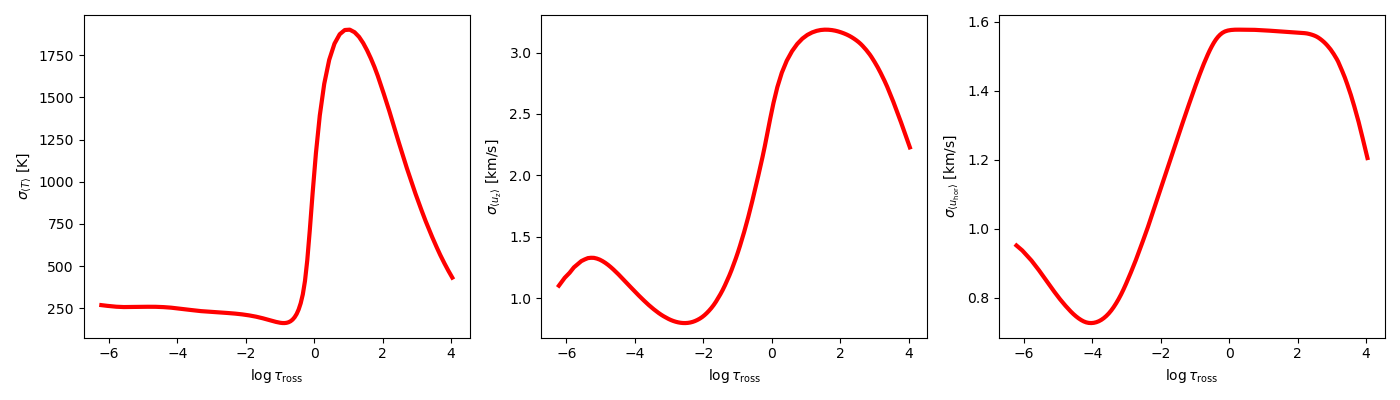}
    \caption{Amplitude of fluctuations in temperature, vertical and horizontal velocity for the Solar \texttt{MURaM} simulation}
    \label{fig:MURAM}
\end{figure*}

Figure \ref{fig:MURAM} shows the temporal fluctuations in temperature, vertical and horizontal velocity for the Solar \texttt{MURaM} simulation used in this work. Specifically, the hydrodynamic quantities were first horizontally averaged for each snapshot in time, and then used to compute the temporal rms standard deviation at each depth point. Take $X$ as the hydrodynamic variable, $\langle X(d)\rangle$ the horizontally and temporally averaged variable, $d$ as the depth variable and $N_\mathrm{snaps}$ as the number of snapshots in the timeseries. The rms is then computed as:

\begin{equation}
    \sigma_X(d) =  \sqrt{\frac{1}{N_\mathrm{snaps}}\sum_{t}(X(d,t)-\langle X(d)\rangle)^2}~~.
\end{equation}

\section{Variability with relation to the line curve-of-growth}
\begin{figure*}
    \centering
    \includegraphics[width = \textwidth, trim=0cm 0cm 0cm 0cm, clip]{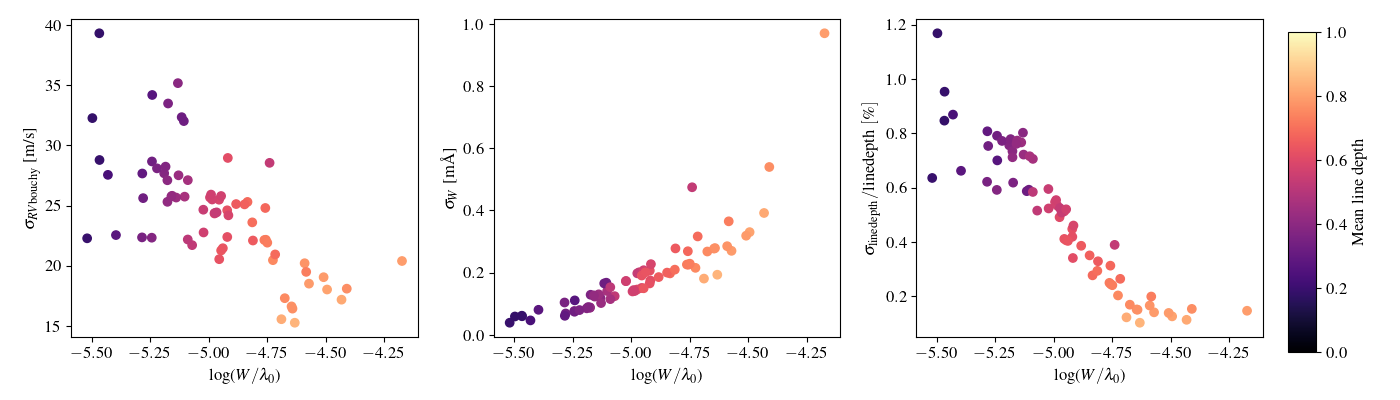}
    \caption{RMS of the temporal evolution of $RV_\mathrm{Bouchy}$, equivalent width and relative line depth at disc centre, for reconstructed lines in the wavelength region $5500-5600$~\AA~ and $6100-6200$~\AA~. All data is plotted relative to reduced equivalent width on the x-axis and colour coded by mean line depth.}
    \label{fig:tempvar_REW}
\end{figure*}

The reduced equivalent width, $W/\lambda_0$, is a proxy of where a line falls on its curve-of-growth. The transition from the weak to the saturated part of the curve-of-growth is element and line dependent. For example for \ion{Fe}{I} this transition is shown to happen around $W/\lambda_0\approx-5$ \citep{Rutten1983,Tielens1999}, where saturated lines have values greater than this transition. The reduced equivalent width is a better proxy for line strength than the line depth, as it is an actual measure of how much radiation is absorbed by the spectral line. The temporal variability of RV, equivalent with and relative line depth, as a function of mean reduced equivalent width is shown in Figure \ref{fig:tempvar_REW}. Regarding $\sigma_{RVbouchy}$ and $\sigma_{W}$, we observe the same monotonic relations as when plotting the variability against mean line depth. Specifically for $\sigma_{W}$, the scatter is reduced and the strongest lines clearly show the largest variability. In the rightmost panel we have plotted the variability in line depth normalised by the mean line depth. It is now clear that the strongest lines exhibit the smallest relative variability in line depth, while weak lines have the largest relative variability.

\section{Wasserstein distances}

\begin{figure*}
    \centering
    \includegraphics[width = \textwidth, trim=0cm 0cm 0cm 0cm, clip]{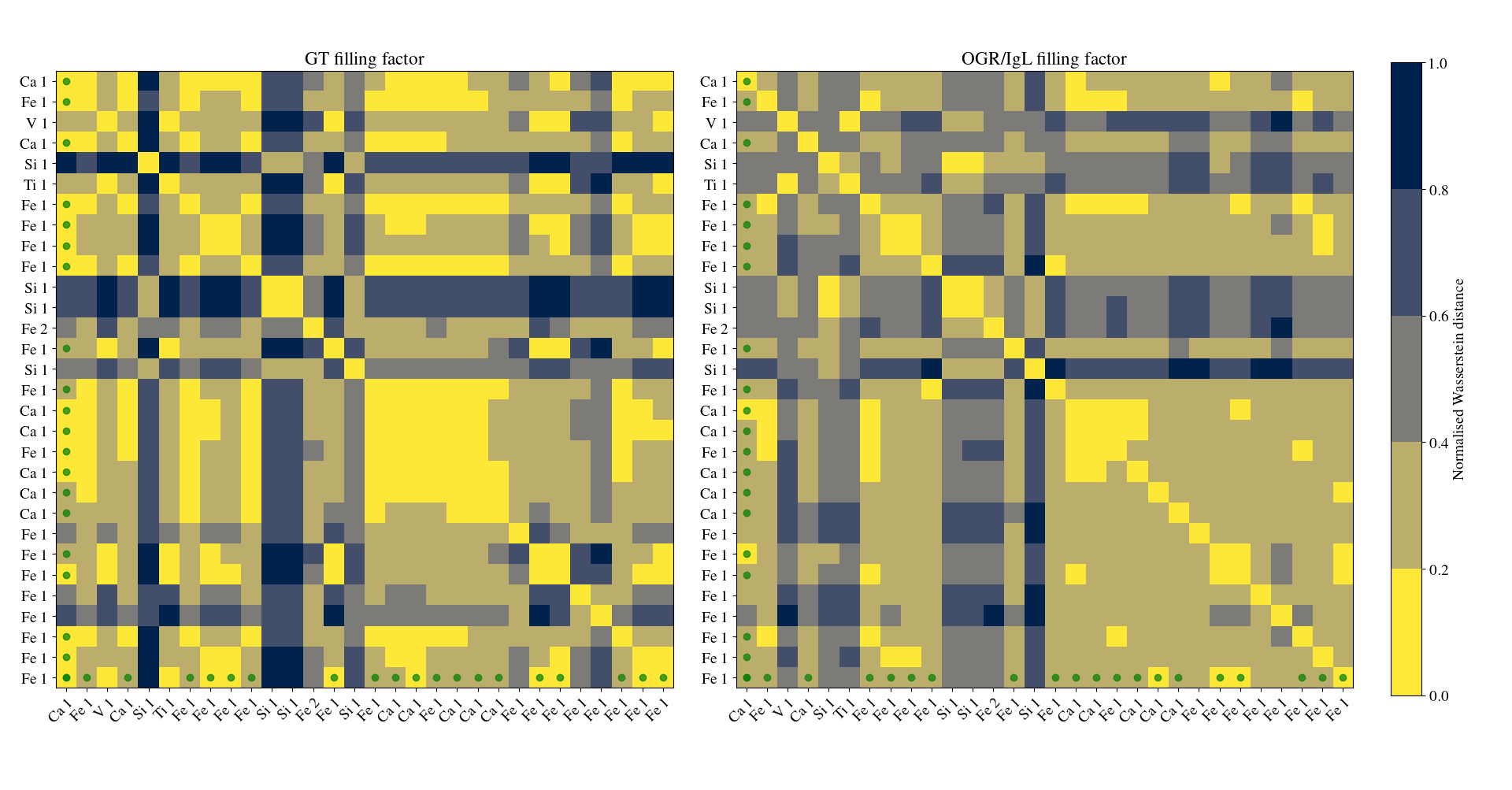}

    \caption{Normalised filling factors for spectral lines in the wavelength region $6100-6200$~\AA~at disc centre. The left and right panel show the 1-Wasserstein distance for the GT and OGR/IgL filling factor distributions, respectively. The green dots visualise the lines that were ultimately selected as a subsample.}
    \label{fig:Wasserstein}
\end{figure*}

Figure \ref{fig:Wasserstein} show the 2D symmetric matrices of 1-Wasserstein distance, for the wavelength region $6100-6200$~\AA~at disc centre, used to select a subsample of similar behaving spectral lines. At each point in the matrix, the 1-Wasserstein distance shows how similar the filling factor distribution of those two spectral lines are. A value of zero means that the two distributions are exactly equal, which is the case on the diagonal. A subsample of similar behaving spectral lines was defined from those lines having a normalised 1-Wasserstein distance less than $0.4$, for both distributions. The final subsample is visualised by green dots, and consists of only \ion{Fe}{I} and \ion{Ca}{I} lines.

% Don't change these lines
\bsp	% typesetting comment
\label{lastpage}
\end{document}